\documentclass[aps,prd,superscriptaddress,showpacs,floatfix,twocolumn,preprintnumbers,nofootinbib]{revtex4}

\usepackage{graphicx}
\usepackage{dcolumn}
\usepackage{amsmath,amssymb,bm}
\usepackage{citesort}
\usepackage{slashed}

\newcommand{\XX}{\mathcal{X}}
\newcommand{\YY}{\mathcal{Y}}
\newcommand{\Order}{\mathcal{O}}
\newcommand{\F}{\mathcal{F}}
\newcommand{\Fpi}{F_\pi}
\newcommand{\Lagr}{\mathcal{L}}

\newcommand{\artanh}{\operatorname{artanh}}
\newcommand{\IR}{\textrm{IR}}
\newcommand{\HB}{\textrm{HB}}

\newcommand{\beq}{\begin{equation}}
\newcommand{\eeq}{\end{equation}}

\newcommand{\nn}{\nonumber\\}

\newcommand{\krig}[1]{\stackrel{\circ}{#1}}

\allowdisplaybreaks[1]

\begin{document}

\preprint{HISKP-TH-09/12, FZJ-IKP-TH-2009-10}

\title{Aspects of meson--baryon scattering in three- and two-flavor\\ chiral perturbation theory}

\author{Maxim Mai}
\email{mai@hiskp.uni-bonn.de}
\affiliation{Universit\"at Bonn,
             Helmholtz-Institut f\"ur Strahlen- und Kernphysik (Theorie)
         and Bethe Center for Theoretical Physics,
             D-53115 Bonn, Germany}

\author{Peter C. Bruns}
\email{bruns@hiskp.uni-bonn.de}
\affiliation{Universit\"at Bonn,
             Helmholtz-Institut f\"ur Strahlen- und Kernphysik (Theorie)
         and Bethe Center for Theoretical Physics,
             D-53115 Bonn, Germany}

\author{Bastian Kubis}
\email{kubis@hiskp.uni-bonn.de}
\affiliation{Universit\"at Bonn,
             Helmholtz-Institut f\"ur Strahlen- und Kernphysik (Theorie)
         and Bethe Center for Theoretical Physics,
             D-53115 Bonn, Germany}

\author{Ulf-G. Mei{\ss}ner}
\email{meissner@hiskp.uni-bonn.de}
\affiliation{Universit\"at Bonn,
             Helmholtz-Institut f\"ur Strahlen- und Kernphysik (Theorie)
         and Bethe Center for Theoretical Physics,
             D-53115 Bonn, Germany}
\affiliation{Forschungszentrum J\"ulich, 
             Institut f\"ur Kernphysik,
             Institute for Advanced Simulation, 
         and J\"ulich Center for Hadron Physics,
             D-52425 J\"ulich, Germany}

\date{\today}

\begin{abstract}
We analyze meson--baryon scattering lengths in the framework of covariant
baryon chiral perturbation theory at leading one-loop order. We 
compute the complete set of matching relations between the dimension-two
low-energy constants in the two- and three-flavor formulations of the theory.
We derive new two-flavor low-energy theorems for pion--hyperon scattering that can be tested in lattice simulations.
\end{abstract}


\maketitle

\section{Introduction and summary}

At energies well below the chiral symmetry breaking scale, chiral
perturbation theory (ChPT) is a useful tool to investigate various 
hadronic interactions, such as meson--baryon scattering (throughout this
paper, mesons and baryons refer to the octet of Goldstone bosons and the
lowest-lying baryon octet, respectively). 
The theory relies on
an expansion of the QCD S-matrix and transition currents in powers of
small external momenta and quark masses; the organization of the different orders of 
any calculation exploits the power counting in these parameters.
[For a recent review on baryon ChPT, see Ref.~\cite{Bernard:2007zu}.]
However, especially for meson--baryon systems, calculations in the three-flavor 
sector of up, down, and strange quarks are often hampered by a slow
convergence due to the large kaon-loop contributions or even require a
nonperturbative resummation due to the appearance of subthreshold resonances.
In Ref.~\cite{Liu:2006xja}
an analysis of meson--baryon scattering lengths at third order in the framework
of heavy-baryon chiral perturbation theory (HBChPT) was presented
(see also Ref.~\cite{Kaiser:2001hr} for an earlier calculation of
kaon--nucleon scattering in that framework). 
This issue may be readdressed in the context of
a covariant formulation of baryon ChPT, as the resummation of the kinetic 
energy terms might help improve the convergence.  We will thus repeat the
calculation of Ref.~\cite{Liu:2006xja} in the framework of 
infrared regularization (IR)~\cite{Becher:1999he}, which also allows to take
the strict heavy-baryon limit. This gives us the opportunity to investigate the question of
how strongly the convergence of the chiral expansion is affected by the heavy-baryon 
limit in meson--baryon scattering at threshold.  

On the other hand, integrating out the strange quarks from the effective
field theory leaves one with  pion--nucleon ChPT, which in general shows
a better convergence due to the much smaller pion mass. Evidently, the two-
and three-flavor theories are not independent -- the pertinent low-energy 
constants (LECs) in the corresponding chiral effective Lagrangian are related by
{\it matching}. As there are more LECs in SU(3) than in SU(2), matching allows
to construct constraints for combinations of three-flavor LECs. Such
constraints can e.g.\ be used in the unitarized coupled-channel 
analysis of the existing and upcoming photo-kaon threshold production data from 
ELSA, JLab, and MAMI~\cite{Borasoy:2007ku}. 
As meson--baryon scattering is an important ingredient for photoproduction
calculations because of final-state interactions, a general matching analysis 
for the LECs related to meson--baryon and pion--nucleon scattering is called
for. This indeed is one of the topics of this paper. 
Furthermore, it was already stressed in the
context of pion--kaon scattering that there exist certain observables in the
three-flavor effective theory that are not afflicted by large kaon-mass corrections~\cite{Roessl}.
We therefore systematically analyze two-flavor versions of pion--hyperon
scattering, extending earlier work on two-flavor ChPT for hyperons~\cite{Beane:2003yx,Tiburzi:2008bk,Jiang:2009sf}.
Such a framework allows to formulate low-energy theorems, which are useful in the
analysis of lattice QCD data~\cite{Torok}.

The pertinent results of our investigation can be summarized as follows:
\begin{itemize}
\item[(i)] We have calculated all elastic meson--baryon scattering lengths in a
  covariant formulation of baryon ChPT to third order in the chiral
  expansion. The appearing dimension-two LECs  were determined from the baryon 
  masses and meson--baryon scattering data, while the dimension-three LECs were set to zero.
  In most channels, we find sizeable second- and third-order corrections to the
  leading-order results.

\item[(ii)] Fitting the dynamical dimension-two LECs to the same input, we find
  that the chiral corrections in the covariant (full) approach come out mostly
  larger than in the heavy-baryon scheme. 

\item[(iii)] We have also considered the so-called reordering prescription~\cite{Becher:1999he} 
  that has proven to be useful 
  in case of the baryon masses and the magnetic moments~\cite{Mojzis:1999qw}. In contrast
  to these quantities, in most cases reordering -- using only the dimension-two 
  operators -- does not improve  the  convergence of the chiral expansion. 

\item[(iv)] We have performed the matching of the two- and three-flavor 
  pion--nucleon scattering amplitudes and derived the complete set 
  of constraints between the dimension-two LECs $c_i$ for SU(2) and 
  $b_i$ for SU(3) up-to-and-including terms of order $M_K \sim \sqrt{m_s}$
.
\item[(v)] We have given the complete effective Lagrangians for pion--hyperon
   interactions to second order in the chiral expansion and 
  derived the matching relations between the dimension-two LECs of the two- and
  three-flavor effective field theories.

\item[(vi)] We have derived novel low-energy theorems for pion--hyperon 
  scattering
  and identified all channels that only acquire corrections in powers of the
  pion mass. These formulas will eventually be useful for chiral
  extrapolation of pion--hyperon scattering from lattice simulations.
\end{itemize}

The article is organized as follows. Section~\ref{sec:MBscatt} contains
the calculation and analysis of the meson--baryon scattering lengths to third 
order in covariant baryon ChPT and the comparison with the earlier heavy-baryon 
results. We also perform reordering of the chiral expansion for the
scattering lengths and discuss the pertinent results. In Sec.~\ref{sec:match},
we consider the matching between the two- and three-flavor versions of the
effective field theory and derive relations between the pertinent dimension-two couplings. 
We derive the general chiral effective Lagrangians for pion--hyperon 
interactions to second order and compute the matching relations to the
three-flavor dimension-two LECs. We analyze the resulting scattering
amplitudes and present new SU(2) low-energy theorems for certain scattering 
amplitudes. We further discuss the resulting chiral extrapolations for all
pion--baryon scattering amplitudes that are not affected by large kaon-loop 
corrections. Most technical details are
relegated to the appendices.

\section{Meson--baryon scattering to one loop}
\label{sec:MBscatt}

In this section we present our results for the S-wave elastic meson--baryon scattering lengths 
up to third chiral order in the framework of
SU(3)  chiral perturbation theory, regularized in a Lorentz-invariant way.

\subsection{Kinematics}

\begin{figure}
\begin{center}\includegraphics[width=0.54\linewidth]{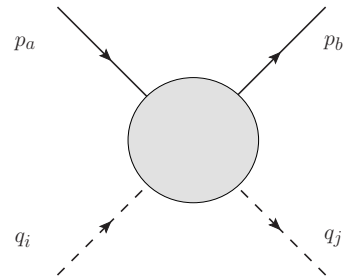}\end{center}
\caption{Kinematical structure of the meson--baryon scattering process.}\label{pic:kinematics}
\end{figure}

In our description of the meson--baryon scattering process we denote the in-
and outgoing momenta as depicted in Fig.~\ref{pic:kinematics}. The 
external particles are on their respective mass shell, 
i.e.\ $p_b^2=p_a^2=m_0^2$ and $q_i^2=q_j^ 2=M_j^2$. 
We neglect isospin breaking and set $m_u=m_d$ such that three different 
meson masses can appear, i.e.\ $\{M_\pi,M_K,M_\eta\}$. 
Since the chiral expansion of the baryon mass goes like
$m_B=m_0+\Order(M_\phi^2)$ 
(here, $\phi$ is a generic symbol for any Goldstone boson)
we are allowed to use a single baryon mass, i.e.\ $m_0$, at the order of the 
calculation considered.

Because of Lorentz invariance and for $p:=p_a+q_i=p_b+q_j$, the on-shell 
scattering amplitude can be decomposed as
\begin{equation}
T^{b,j,i,a}=T_1^{b,j,i,a}+\slashed{p} \, T_2^{b,j,i,a}~,\label{eqn:T}
\end{equation}
with $a,b$ ($i,j$) baryon (meson) flavor indices.
The scattering amplitude at threshold, to which we restrict our
calculation, determines the S-wave scattering length
\begin{equation}
a_{\phi B}^{b,j,i,a}=\frac{m_B}{4\pi(m_B+M_j)}T_{\phi B}^{b,j,i,a} ~,\label{eqn:kinematics}
\end{equation}
where $T_{\phi B}^{b,j,i,a}$ denotes the S-wave projection of the full
scattering amplitude, given in~\eqref{eqn:T}, at threshold.
Note that here $m_B\in\{m_N,m_\Lambda,m_\Sigma,m_\Xi\}$, i.e.\ we use
the physical masses for the normalization of the scattering lengths;
this is consistent within the power counting scheme employed.
Assuming isospin invariance and applying the Wigner--Eckhart theorem, the
minimal set of  T-matrices reduces to $26$ different combinations, 
which fully determine  meson--baryon scattering. 
For total isospin~$I$ of the meson--baryon system, we denote its elements 
by $T_{\phi B}^{({I})}$ and the corresponding scattering lengths by 
$a_{\phi B}^{({I})}$.

\subsection{Effective Lagrangian}

For the calculation of the scattering amplitudes up-to-and-including the third 
chiral order one requires the explicit form of the chiral  Lagrangian
\begin{equation}
\Lagr=\Lagr^{(2)}_{\phi}+\Lagr^{(4)}_{\phi}+\Lagr^{(1)}_{\phi
  B}+\Lagr^{(2)}_{\phi B}+\Lagr^{(3)}_{\phi B}~.
\end{equation}
The relevant degrees of freedom are the Goldstone bosons described 
by the traceless meson matrix $U\in SU(3)$,
\begin{equation}\label{eqn:fields}
U =\exp\Bigl(i\frac{\phi}{F_0}\Bigr), ~ \phi=\sqrt{2}\begin{pmatrix}
\frac{\pi^0}{\sqrt{2}}+\frac{\eta}{\sqrt{6}} \!\!&\!\! \pi^+ \!\!&\!\! K^+ \\
\pi^- \!\!&\!\! -\frac{\pi^0}{\sqrt{2}}+\frac{\eta}{\sqrt{6}} \!\!&\!\! K^0 \\
K^- \!\!&\!\! \bar{K}^0 \!\!&\!\! -\frac{2}{\sqrt{6}}\eta
\end{pmatrix},
\end{equation}
where $F_0$ is the meson decay constant in the chiral limit, and the
low-lying baryons, collected in a traceless matrix
\begin{eqnarray}\label{eqn:baryonmatrix}
B=\begin{pmatrix}
\frac{\Sigma^0}{\sqrt{2}}+\frac{\Lambda}{\sqrt{6}} & \Sigma^+ & p \\
\Sigma^- & -\frac{\Sigma^0}{\sqrt{2}}+\frac{\Lambda}{\sqrt{6}}& n \\
\Xi^-& {\Xi}^0 & -\frac{2}{\sqrt{6}}\Lambda
\end{pmatrix}~.
\end{eqnarray}
We set external currents to zero except the scalar $s$, which contains  
the quark mass matrix 
$s= \mathcal{M}=\textrm{diag}(\hat m,\hat m, m_s)$,
where $\hat m$ is the average light quark mass. We furthermore use
\begin{align}
u^2:=U  ~,\quad
u^\mu:=iu^{\dagger}\partial^\mu u +iu\partial^\mu u^{\dagger} ~,\nn
\quad
\chi_\pm:=u^{\dagger}\chi u^{\dagger}\pm u\chi^{\dagger}u ~,\quad
\chi:=2B_0\, s ~,
\end{align}
where $B_0$ is a constant related to the quark condensate in the chiral limit.
The leading-order Lagrangian reads
\begin{align}
\Lagr^{(1)}_{\phi B}&=\langle \bar{B} (i\gamma_\mu D^\mu-m_0)B\rangle
+\frac{D/F}{2}\langle \bar{B}\gamma_\mu \gamma_5[u^\mu,B]_\pm \rangle ~,
\end{align}
where $\langle\ldots\rangle$ denotes the trace in flavor space,
$D_\mu B :=\partial_\mu B +\frac{1}{2}[[u^\dagger,\partial_\mu u],B]$, 
$m_0$ is the common baryon octet mass in the chiral limit,  
and $D$, $F$ are the axial coupling constants.

The next-to-leading-order Lagrangian~\cite{Krause:1990xc} in its 
minimal form contains, for the specific case of meson--baryon
scattering, 14 independent structures~\cite{Frink:2005ru}, 
where the first three express the explicit symmetry breaking through the 
nonvanishing quark masses:
\begin{align}
\Lagr^{(2)}_{\phi B}&= b_{D/F} \langle\bar B \left[\chi_+,B\right]_\pm\rangle+b_0 \langle\bar B B\rangle \langle\chi_+\rangle\nn
&+b_{1/2} \langle\bar B  \left[u_\mu,\left[u^\mu,B\right]_\mp\right]\rangle \nn
&+b_3 \langle\bar B \left\{ u_\mu,\left\{ u^\mu,B\right\}\right\}\rangle
+b_4 \langle\bar B  B\rangle \langle u_\mu u^\mu\rangle \nn
&+i\sigma^{\mu\nu}\Big(b_{5/6}  \langle\bar B \left[\left[u_\mu,u_\nu\right], B\right]_\mp\rangle
+b_7 \langle\bar Bu_\mu\rangle  \langle u_\nu B\rangle\Big) \nn
&+ \frac{i\,b_{8/9}}{2m_0}\Big( \langle\bar B \gamma^\mu\left[u_\mu,\left[u_\nu,\left[D^\nu, B\right]\right]_\mp\right]\rangle\nn
&\qquad\quad+\langle\bar B \gamma^\mu\big[D_\nu,\left[u^\nu,\left[u_\mu,B\right]\right]_\mp\big]\rangle\Big) \nn
&+\frac{i\,b_{10}}{2m_0}\Big( \langle\bar B \gamma^\mu\left\{ u_\mu,\left\{ u_\nu,\left[D^\nu,B\right]\right\}\right\}\rangle\nn
&\qquad\quad+\langle\bar B\gamma^\mu\left[D_\nu,\left\{ u^\nu,\left\{ u_\mu,B\right\}\right\}\right]\rangle\Big) \nn
&+\frac{i\,b_{11}}{2m_0}\Big( 2\langle\bar B \gamma^\mu \left[D_\nu,B\right]\rangle \langle u_\mu u^\nu\rangle\nn
&\qquad\quad+\langle\bar B \gamma^\mu B\rangle \langle\left[D_\nu,u_\mu\right]u^\nu + u_\mu \left[D_\nu,u^\nu\right]\rangle   \Big)~,\label{eq:L2}
\end{align}
with the $b_i$ the pertinent dimension-two LECs. The
LECs $b_{0,D,F}$ are the so-called {\it symmetry breakers} while the $b_i$ $(i= 1,
\ldots, 11)$ are the {\it dynamical} LECs.

At next-to-next-to-leading order the number of independent structures
increases to 78~\cite{Frink:2006hx} (see also Ref.~\cite{Oller:2007qd}).
In Ref.~\cite{Liu:2006xja}, the finite contributions from these 
dimension-three LECs were omitted.
We will follow that approach here for the purpose of a direct comparison with the 
heavy-baryon results~\cite{Liu:2006xja},
and therefore refrain from showing the third-order counterterms
explicitly. 
We remark that the matching relations for the dimension-two LECs at next-to-leading order in Sec.~\ref{sec:match}
are not affected by any of the dimension-three terms.

We will also require the purely mesonic Lagrangian at leading and
next-to-leading order
\begin{align}
\Lagr^{(2)}_\phi+\Lagr^{(4)}_\phi&= \frac{F_0^2}{4}\langle u_\mu u^\mu  +\chi_+\rangle 
\nn
&+L_4\langle u_\mu u^\mu \rangle\langle\chi_+\rangle+L_5\langle u_\mu u^\mu \chi_+ \rangle + \ldots ~,
\end{align}
where again we have only displayed the terms required later.

We wish to remark here that we stick to baryon chiral perturbation theory \emph{without}
explicit decuplet contributions, for the following reasons.
For the observables under investigation in the present article, i.e.\ S-wave 
scattering lengths, the decuplet does not contribute at tree level.
Furthermore, the effects of the remaining $\pi\Delta$ loop contributions have been shown
to be small for the $\pi N$ scattering lengths in a corresponding SU(2) 
calculation~\cite{Fettes:2000bb}.
Decuplet contributions to meson--baryon scattering in the heavy-baryon formalism
have been investigated in Ref.~\cite{Liu:2007ct}.

\subsection{Diagrammatics}

\begin{figure}
\begin{center}\includegraphics[width=0.7\linewidth]{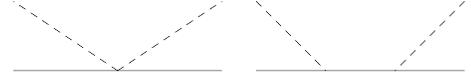}\end{center}
\caption{Weinberg--Tomozawa (left) and Born (right) type diagrams. Crossed graphs are not shown.}\label{pic:WT-BORN}
\end{figure}
\begin{figure*}
\begin{center}\includegraphics[width=0.6\linewidth]{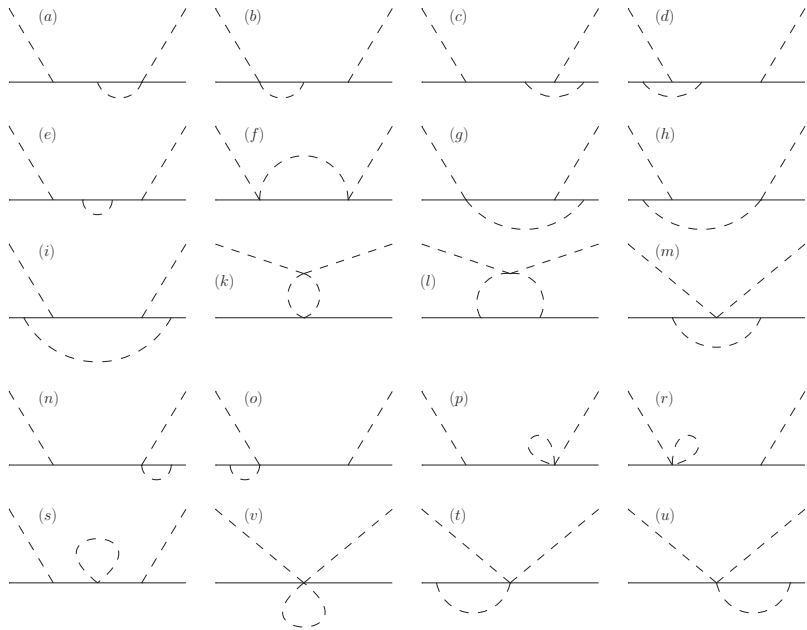}
\caption{One-loop contributions to $\phi B$ scattering without external leg corrections and crossed graphs. 
Graphs containing closed fermion loops are not shown either, since they vanish in infrared regularization.}
\label{pic:graphs}
\end{center}
\end{figure*}
\begin{figure}
\begin{center}\includegraphics[width=\linewidth]{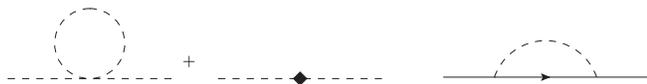}
\caption{Wave-function renormalization contributions for mesons (dashed) 
and baryons (solid). 
Counterterm insertions from  $\Lagr_{\phi}^{(4)}$ are denoted by the diamond.}\label{pic:Z}
\end{center}
\end{figure}

At the first chiral order the scattering amplitudes are determined by the
contact terms from $\Lagr^{(1)}_{\phi B}$, i.e.\ the so-called
Weinberg--Tomozawa term, as well as Born graphs, see Fig.~\ref{pic:WT-BORN}. 
To the second chiral order only contact terms of $\Lagr^{(2)}_{\phi B}$ contribute.
The third-order scattering amplitude contains both contact terms from
$\Lagr^{(3)}_{\phi B}$ and loop diagrams, see Fig.~\ref{pic:graphs}
(note that the tree graphs are not shown because the corresponding LECs are set to zero).
In addition, wave-function renormalization of the external baryon and meson legs 
must be taken into account, which is 
determined by the self-energy diagrams in Fig.~\ref{pic:Z}.

\subsection{Infrared regularization}

Loop corrections are in general divergent and require
regularization and renormalization. In the baryon sector, a straightforward
application of dimensional regularization in the covariant theory is not possible 
as the large scale related to the baryon mass in the chiral limit invalidates
the power counting~\cite{Gasser:1987rb}.
This problem can be overcome utilizing the  so-called infrared regularization as
developed in Ref.~\cite{Becher:1999he} (see also Ref.~\cite{EllisTang}). 
In essence, any one-loop integral can be split into
an infrared-singular and a -regular part by rearranging the Feynman parameter
integration (we consider here, as an example, a loop diagram with one meson and one baryon propagator)
\begin{equation}\label{eq:split}
\int_0^1dz \, (\ldots)=\int_0^\infty dz \, (\ldots)-\int_1^\infty dz\, (\ldots) ~.
\end{equation}
The infrared singularity is generated in the region of small values of the parameter
$z$, so that this term contains all the relevant low-energy information.
On the other hand, the second integral on the right-hand side of~\eqref{eq:split} 
can be expanded in $M/m$ as (as always, $M \ (m)$ denotes a meson (baryon) mass)
\begin{equation}
m^{d-4}\, \Big( d_0+d_1\Big(\frac{M}{m}\Big)+ d_2\Big(\frac{M}{m}\Big)^2+\ldots\Big)\,,
\end{equation}
which is called the regular part of the loop integral. Its contribution can be
absorbed in the LECs of the effective Lagrangian and thus does not need to be
considered explicitly. The remaining  infrared-singular part of 
the loop integral has a different low-energy expansion,
\begin{equation}
m^{d-4} \, \Big(\frac{M}{m}\Big)^{d-3} \, \Big( c_0+c_1\Big(\frac{M}{m}\Big)
+ c_2\Big(\frac{M}{m}\Big)^2+\ldots\Big)~,
\end{equation}
and thus it is proportional to a $d$-dependent (fractional) power of $M/m$. 
By the means of analytic continuation we can manipulate any one-loop integral 
such that the limit $d=4$ can be performed. Notice that the dimensionality of the Feynman
parameter space grows with increasing number of propagators. 
We furthermore remark that loop integrals consisting solely of meson propagators
are unchanged compared to dimensional regularization, while those containing exclusively
baryon propagators have no infrared-singular part and vanish in this scheme.
For a more complete treatment of such integrals we refer to the original
article~\cite{Becher:1999he}.

Within our calculation we reduce all loop integrals to 
linear combinations of scalar loop integrals by the 
Passarino--Veltman reduction procedure~\cite{Passarino:1978jh}, 
which can be evaluated in the sense of the above discussion. 
However, since there is no explicit representation of scalar loop integrals
with more than one baryon propagator in terms of elementary functions, we evaluate the
scalar loop integrals in a small region around threshold, see Appendix~\ref{App:thr-exp},
where we obtain elementary functions of the meson and baryon masses, 
see Appendix~\ref{App:Integrals-results}.  
We check our final result by expanding it to a finite chiral order, obtaining the heavy-baryon result for
meson--baryon scattering~\cite{Kaiser:2001hr,Liu:2006xja}. 
Moreover, we have checked that after a projection to 
the SU(2) subgroup of SU(3) our result is consistent with the known
IR-representation of the pion--nucleon scattering amplitude~\cite{Becher:2001hv}.

\subsection{Results and discussion}
\subsubsection{Formal results}

At first and second order in the chiral expansion, the T-matrix elements are determined by tree graphs
with vertices from $\Lagr_{\phi B}^{(1)}$ and $\Lagr_{\phi B}^{(2)}$, and thus are
independent of any regularization scheme.
Born type graphs contribute to the first chiral order of the scattering amplitude, 
however at threshold their leading contribution is shifted to second order.  
We still count these terms as part of the first-order corrections. 
This is in contrast to the discussions in HBChPT~\cite{Kaiser:2001hr,Liu:2006xja}, 
where strict $1/m_0$ expansion applies and Born term contributions are part of the second order.

At second order only four independent combinations $\{B_1,B_2,B_3,B_4\}$
of the dynamical LECs $\{b_1,\ldots,b_{11}\}$ of 
$\Lagr_{\phi B}^{(2)}$ appear in meson--baryon threshold amplitudes,
\beq
B_1= b_1+b_8\,,~ B_2= b_2+b_9\,,~ B_3= b_3+b_{10}\,,~ B_4= b_4+b_{11}\,,
\label{eqn:mapping}
\eeq
for which we will quote fit results below.

To third chiral order both one-loop diagrams as well as contact terms from 
$\Lagr_{\phi B}^{(3)}$ contribute. 
As there is no sufficient experimental information to fix the latter,
we restrict ourselves to the one-loop corrections at this order (consistent
with the heavy-baryon calculation~\cite{Liu:2006xja}).
Here as well as in the first- and second-order contributions we replace 
the decay constants in the chiral limit by the physical values corresponding 
to the chosen channel. 
Since the chiral expansion of the decay constants~\cite{GL:NPB250} is of the form 
$F_{\pi,K,\eta}=F_0+\Order(q^2)$, the latter replacement within the
second- and third-order corrections only changes
the final result at orders beyond the accuracy considered here. 
The replacement of the decay constants in the leading-order
contributions to the meson--baryon scattering amplitudes produces a correction
to the third order, which must be taken into account.

Unfortunately the results of the \textit{full} scattering amplitudes become
rather lengthy.\footnote{These results can be obtained as a $Mathematica^{\text{\textregistered}}$.nb-file 
on request to \texttt{mai@hiskp.uni-bonn.de}}
The far more compact \textit{truncated} result (here and in
what follows, we refer to the heavy-baryon calculation
to third order in the chiral expansion as the truncated result), where we 
neglect all contributions starting with the fourth chiral order in the 
\textit{full} one-loop contributions, agrees completely with those 
from Refs.~\cite{Kaiser:2001hr,Liu:2006xja}.

\subsubsection{Fitting of the LECs}\label{sec:fits}

We use the average octet baryon mass $m_0=1.150$~GeV as well as the following numerical input:
$M_\pi = 139.57$~MeV, $M_K = 493.68$~MeV, $M_\eta = 547.75$~MeV, $F_\pi = 92.4$~MeV,
 $F_K =113.0$~MeV, $F_\eta = 120.1$~MeV, $D =0.8$, and $F =0.5$.
We vary the scale $\mu$ in a reasonable range around $m_0$, 
$0.938\text{ GeV}<\mu<1.314\text{ GeV}$.

As already stressed for a direct comparison with the HBChPT results of 
Ref.~\cite{Liu:2006xja}, we will neglect the third-order contact terms.  
Still, we must determine the  second-order LECs, 
i.e.\ $\{b_0$, $b_F$, $b_D$, $B_1$, $B_2$, $B_3$, $B_4\}$. 
These can be obtained by fitting the scattering amplitudes to experimental
results for pion--nucleon and kaon--nucleon scattering, 
for which we use, in agreement with the analysis of Ref.~\cite{Liu:2006xja}, 
$a_{\pi N}^{+}:=(2a_{\pi N}^{3/2}+a_{\pi N}^{1/2})/3=-0.002\pm0.007$~fm \cite{Schroeder} and $a_{K N}^{(0)}=+0.02$~fm 
as well as $a_{K N}^{(1)}=-0.33$~fm \cite{Martin:1980qe}.

\begin{table*}
\begin{center}
\renewcommand{\arraystretch}{1.5}
\begin{tabular}{rcclll}
\hline\hline
Channel&$=$&$\Order(q^1) $&$+\Order(q^2)$&$+\Order(q^3)_{IR}$&$\qquad\sum_{IR}$\\
\hline
$a_{\pi N}^{(3/2)}$&$=$&$\qquad-0.12\qquad$&$+0.05_{-0.06}^{+0.06}\qquad$&$+0.04_{-0.01}^{+0.01}\qquad$&$-0.04_{-0.07}^{+0.07}$\\
$a_{\pi N}^{(1/2)}$&$=$&$+0.21$&$+0.05_{-0.06}^{+0.06}$&$-0.19_{-0.01}^{+0.01}$&$+0.07_{-0.07}^{+0.07}$\\
$a_{\pi \Xi}^{(3/2)}$&$=$&$-0.12$&$+0.04_{-0.07}^{+0.06}$&$+0.10_{+0.00}^{+0.00}$&$+0.02_{-0.07}^{+0.06}$\\
$a_{\pi \Xi}^{(1/2)}$&$=$&$+0.23$&$+0.04_{-0.07}^{+0.06}$&$-0.24_{-0.03}^{+0.02}$&$+0.02_{-0.10}^{+0.08}$\\
$a_{\pi \Sigma}^{(2)}$&$=$&$-0.24$&$+0.10_{-0.03}^{+0.02}$&$+0.15_{-0.01}^{+0.02}$&$+0.01_{-0.04}^{+0.04}$\\
$a_{\pi \Sigma}^{(1)}$&$=$&$+0.22$&$+0.09_{-0.15}^{+0.15}$&$-0.21_{-0.02}^{+0.01}$&$+0.10_{-0.17}^{+0.16}$\\
$a_{\pi \Sigma}^{(0)}$&$=$&$+0.46$&$+0.11_{-0.17}^{+0.15}$&$-0.47_{-0.03}^{+0.02}$&$+0.10_{-0.19}^{+0.17}$\\
$a_{\pi \Lambda}^{(1/2)}$&$=$&$-0.01$&$+0.03_{-0.03}^{+0.03}$&$-0.03_{-0.01}^{+0.01}$&$-0.01_{-0.04}^{+0.04}$\\
\hline
$a_{K N}^{(1)}  $&$=$&$-0.45$&$+0.60_{-0.20}^{+0.14}$&$-0.48_{-0.12}^{+0.18}$&$-0.33_{-0.32}^{+0.32}$\\
$a_{K N}^{(0)}  $&$=$&$+0.04$&$-0.15_{-0.61}^{+0.59}$&$+0.13_{-0.03}^{+0.05}$&$+0.02_{-0.64}^{+0.64}$\\
$a_{\bar{K} N}^{(1)} $&$=$&$+0.20$&$+0.22_{-0.40}^{+0.36}$&$-0.26_{-0.03}^{+0.02}+0.18 i\qquad$&$+0.16_{-0.44}^{+0.39}+0.18 i$\\
$a_{\bar{K} N}^{(0)} $&$=$&$+0.53$&$+0.97_{-0.51}^{+0.42}$&$-0.40_{-0.08}^{+0.05}+0.22 i$&$+1.11_{-0.59}^{+0.47}+0.22 i$\\
$a_{K \Sigma}^{(3/2)} $&$=$&$-0.31$&$+0.33_{-0.41}^{+0.41}$&$-0.30_{-0.07}^{+0.11}+0.12 i$&$-0.28_{-0.49}^{+0.52}+0.12 i$\\
$a_{K \Sigma}^{(1/2)} $&$=$&$+0.47$&$+0.19_{-0.57}^{+0.50}$&$+0.20_{-0.07}^{+0.05}+0.01 i$&$+0.87_{-0.64}^{+0.55}+0.01 i$\\
$a_{\bar{K} \Sigma}^{(3/2)}$&$=$&$-0.22$&$+0.24_{-0.44}^{+0.39}$&$-0.35_{-0.03}^{+0.05}+0.08 i$&$-0.33_{-0.47}^{+0.44}+0.08 i$\\
$a_{\bar{K} \Sigma}^{(1/2)}$&$=$&$+0.34$&$+0.38_{-0.52}^{+0.55}$&$+0.27_{-0.06}^{+0.04}+0.01 i$&$+0.98_{-0.59}^{+0.59}+0.01 i$\\
$a_{K \Xi}^{(1)}  $&$=$&$+0.15$&$+0.34_{-0.43}^{+0.43}$&$-0.02_{-0.01}^{+0.00}+0.17 i$&$+0.48_{-0.43}^{+0.43}+0.17 i$\\
$a_{K \Xi}^{(0)}  $&$=$&$+0.66$&$+0.98_{-0.58}^{+0.45}$&$-0.62_{-0.09}^{+0.06}+0.14 i$&$+1.02_{-0.68}^{+0.51}+0.14 i$\\
$a_{\bar{K} \Xi}^{(1)} $&$=$&$-0.50$&$+0.66_{-0.22}^{+0.15}$&$-0.42_{-0.12}^{+0.19}$&$-0.26_{-0.34}^{+0.34}$\\
$a_{\bar{K} \Xi}^{(0)} $&$=$&$-0.15$&$+0.02_{-0.63}^{+0.70}$&$+0.13_{-0.05}^{+0.08}$&$+0.00_{-0.68}^{+0.78}$\\
$a_{K \Lambda}^{(1/2)}$&$=$&$-0.04$&$+0.50_{-0.50}^{+0.46}$&$-0.27_{-0.06}^{+0.10}+0.14 i$&$+0.19_{-0.56}^{+0.55}+0.14 i$\\
$a_{\bar{K} \Lambda}^{(1/2)}$&$=$&$-0.05$&$+0.50_{-0.50}^{+0.46}$&$-0.40_{-0.06}^{+0.09}+0.18 i$&$+0.04_{-0.56}^{+0.55}+0.18 i$\\
\hline
$a_{\eta N}^{(1/2)}$&$=$&$-0.01$&$+0.26_{-0.62}^{+0.56}$&$-0.13_{-0.03}^{+0.04}+0.19 i$&$+0.13_{-0.65}^{+0.60}+0.19 i$\\
$a_{\eta \Xi}^{(1/2)}$&$=$&$-0.09$&$+0.84_{-0.67}^{+0.65}$&$-0.49_{-0.06}^{+0.09}+0.17 i$&$+0.25_{-0.73}^{+0.74}+0.17 i$\\
$a_{\eta \Sigma}^{(1)}$&$=$&$-0.04$&$+0.22_{-0.22}^{+0.20}$&$-0.15_{-0.02}^{+0.03}+0.13 i$&$+0.03_{-0.24}^{+0.24}+0.13 i$\\
$a_{\eta \Lambda}^{(0)}$&$=$&$-0.04$&$+0.70_{-0.47}^{+0.38}$&$-0.51_{-0.08}^{+0.13}+0.38 i$&$+0.15_{-0.55}^{+0.51}+0.38 i$\\
\hline
\hline
\end{tabular}
\renewcommand{\arraystretch}{1.0}
\caption{\textit{Full} result for the fit to $a^{+}_{\pi N}$, $a_{K N}^{(1)}$, $a_{K N}^{(0)}$  
  and $d_0=-0.996$~GeV$^{-1}$. All scattering lengths are given in units of
  fm. $\Sigma$ denotes  the sum of all contributions from first, second, and
  third order. }\label{tab:full-fit}
\end{center}
\end{table*}

Our fitting procedure contains two steps. First we fix the symmetry
breakers $\{b_0$, $b_D$, $b_F\}$ using the formulas for the baryon masses as
well as the formula for the pion--nucleon sigma term $\sigma_{\pi N}$ up to
third order from earlier IR and HBChPT calculations~\cite{Ellis:1999jt,Bernard:1993nj}, 
and fitting the LECs to the physical 
values, $m_N=0.938$~GeV, $m_\Lambda=1.115$~GeV, $m_\Sigma=1.192$~GeV, $m_\Xi=1.314$~GeV, 
and $\sigma_{\pi N}=45\pm 8$~MeV~\cite{Gasser:1990ce}. 
The result reads 
\begin{align}
b^\IR_0&=-0.45_{-0.01}^{+0.01}\text{ GeV}^{-1}~, & 
b^\HB_0&=-0.48\text{ GeV}^{-1} ~, \nn
b^\IR_D&=+0.05_{-0.00}^{+0.00}\text{ GeV}^{-1}~, &
b^\HB_D&=+0.05\text{ GeV}^{-1}~, \nn
b^\IR_F&=-0.45_{-0.01}^{+0.01}\text{ GeV}^{-1}~,&
b^\HB_F&=-0.47\text{ GeV}^{-1}~.
\end{align}
Using this result we fix in the second step the remaining, dynamical,
LECs, by fitting the result for the scattering lengths up to third order to 
the experimental data for $a^+_{\pi N}$, $a_{K N}^{(1)}$, and $a_{K N}^{(0)}$. 
However to fix four independent parameters we need at least one more input,
for which we choose a particular linear combination of heavy-baryon LECs
$d_0=-0.996$ GeV$^{-1}$, in agreement with the discussions of the scattering lengths
within HBChPT~\cite{Kaiser:2001hr,Liu:2006xja}. 
$d_0$ is determined within a coupled-channel approach~\cite{J.CaroRamon} 
and is given by $d_0=-2B_1+2B_3+2B_4$. 
The results for the \textit{full} (IR) and \textit{truncated} (HB) one-loop 
corrections read
\begin{align}
B^\IR_1&=+0.43_{-0.12}^{+0.08}\text{GeV}^{-1}, &  B^\HB_1&=+0.29_{-0.04}^{+0.02}\text{GeV}^{-1},\nn[1mm]
B^\IR_2&=-0.34_{-0.25}^{+0.30}\text{GeV}^{-1}, &  B^\HB_2&=-0.35_{-0.13}^{+0.14}\text{GeV}^{-1},\nn[1mm]
B^\IR_3&=+0.46_{-0.40}^{+0.34}\text{GeV}^{-1}, &  B^\HB_3&=+0.10_{-0.18}^{+0.16}\text{GeV}^{-1},\nn[1mm]
B^\IR_4&=-0.53_{-0.32}^{+0.22}\text{GeV}^{-1}, &  B^\HB_4&=-0.31_{-0.22}^{+0.18}\text{GeV}^{-1}.
\end{align}

The results for the scattering lengths are collected in Tables~\ref{tab:full-fit} and
\ref{tab:truncated-fit}. 
The errors quoted there are of twofold origin: they comprise the uncertainties of the experimental input, 
as well as the change due to the variation of the renormalization scale $\mu$.
The latter effect is pronounced due to the neglect of the dimension-three LECs, whose scale
dependence cancels the one of the loops at third order.  
For the pion channels we find that the
\textit{truncated} results show better convergence as well as consistency
with the experimental data.  However in the kaon channels
both approaches reproduce  $a_{\bar KN}^{(1)}=(+0.37+0.60i)$~fm~\cite{Martin:1980qe} 
quite well.  $a_{\bar KN}^{(0)}$ 
appears with the wrong sign compared to $a_{\bar KN}^{(0)}=(-1.70+0.68i)$~fm~\cite{Martin:1980qe}; 
the failure to describe this particular scattering length  
can be understood by the presence of the subthreshold resonance $\Lambda(1405)$. 
On the other hand, information on the $\eta N$ scattering length, where the
corresponding threshold lies only $\approx 50$~MeV below the $S_{11}$ resonance $N(1535)$,
has been obtained
in a coupled-channel analysis of the inelastic reactions $\pi^-p\rightarrow \eta n$,
with the result $a_{\eta N}^{(1/2)}=\big(+0.621\pm 0.040+(0.306\pm0.034)i\big)$~fm~\cite{Abaev:1996vn}.
The real part is reproduced quite well in both approaches, the imaginary
part deviates slightly from the coupled-channel result. 
Note also that the
small value for the $\pi\Lambda$ scattering length is consistent with the
measured phase shift difference at the $\Xi$ mass~\cite{Chakravorty:2003je,Huang:2004jp}.
In both approaches the chiral series of the scattering lengths in the 
nonpionic channels do not (or at most case by case) converge.

\begin{table*}
\begin{center}
\renewcommand{\arraystretch}{1.5}
\begin{tabular}{rcclll}
\hline
\hline
Channel&$=$&$\Order(q^1) $&$+\Order(q^2)$&$+\Order(q^3)_{\HB}$&$\qquad\sum_{\HB}$\\
\hline
$a_{\pi N}^{(3/2)}$&$=$&$\qquad-0.12\qquad$&$+0.05_{-0.03}^{+0.02}\qquad$&$-0.06_{+0.00}^{+0.00}\qquad$&$-0.13_{-0.03}^{+0.03}$\\
$a_{\pi N}^{(1/2)}$&$=$&$+0.21$&$+0.05_{-0.03}^{+0.02}$&$+0.00_{+0.00}^{+0.00}$&$+0.26_{-0.03}^{+0.03}$\\
$a_{\pi \Xi}^{(3/2)}$&$=$&$-0.12$&$+0.04_{-0.03}^{+0.03}$&$-0.09_{+0.00}^{+0.00}$&$-0.17_{-0.03}^{+0.03}$\\
$a_{\pi \Xi}^{(1/2)}$&$=$&$+0.23$&$+0.04_{-0.03}^{+0.03}$&$-0.03_{+0.00}^{+0.00}$&$+0.23_{-0.03}^{+0.03}$\\
$a_{\pi \Sigma}^{(2)}$&$=$&$-0.24$&$+0.07_{-0.01}^{+0.01}$&$-0.07_{+0.00}^{+0.00}$&$-0.24_{-0.01}^{+0.01}$\\
$a_{\pi \Sigma}^{(1)}$&$=$&$+0.22$&$+0.11_{-0.06}^{+0.06}$&$+0.00_{+0.00}^{+0.00}$&$+0.33_{-0.06}^{+0.06}$\\
$a_{\pi \Sigma}^{(0)}$&$=$&$+0.46$&$-0.01_{-0.07}^{+0.07}$&$+0.04_{-0.01}^{+0.01}$&$+0.49_{-0.08}^{+0.07}$\\
$a_{\pi \Lambda}^{(1/2)}$&$=$&$-0.01$&$+0.03_{-0.01}^{+0.01}$&$-0.11_{+0.00}^{+0.00}$&$-0.09_{-0.01}^{+0.01}$\\
\hline
$a_{K N}^{(1)}  $&$=$&$-0.45$&$+0.40_{-0.06}^{+0.04}$&$-0.28_{-0.04}^{+0.06}$&$-0.33_{-0.10}^{+0.10}$\\
$a_{K N}^{(0)}  $&$=$&$+0.04$&$+0.04_{-0.27}^{+0.27}$&$-0.06_{+0.00}^{+0.00}$&$+0.02_{-0.27}^{+0.27}$\\
$a_{\bar{K} N}^{(1)} $&$=$&$+0.20$&$+0.22_{-0.16}^{+0.15}$&$-0.02_{-0.03}^{+0.02}+0.36 i\qquad$&$+0.40_{-0.19}^{+0.17}+0.36 i$\\
$a_{\bar{K} N}^{(0)} $&$=$&$+0.53$&$+0.58_{-0.20}^{+0.17}$&$+0.41_{-0.09}^{+0.06}+0.22 i$&$+1.52_{-0.29}^{+0.22}+0.22 i$\\
$a_{K \Sigma}^{(3/2)} $&$=$&$-0.31$&$+0.33_{-0.17}^{+0.17}$&$-0.06_{-0.02}^{+0.03}+0.16 i$&$-0.04_{-0.19}^{+0.20}+0.16 i$\\
$a_{K \Sigma}^{(1/2)} $&$=$&$+0.47$&$+0.19_{-0.24}^{+0.22}$&$+0.16_{-0.06}^{+0.04}+0.04 i$&$+0.83_{-0.30}^{+0.26}+0.04 i$\\
$a_{\bar{K} \Sigma}^{(3/2)}$&$=$&$-0.22$&$+0.24_{-0.18}^{+0.16}$&$-0.26_{-0.02}^{+0.03}+0.16 i$&$-0.24_{-0.20}^{+0.20}+0.16 i$\\
$a_{\bar{K} \Sigma}^{(1/2)}$&$=$&$+0.34$&$+0.38_{-0.23}^{+0.23}$&$+0.56_{-0.06}^{+0.04}+0.04 i$&$+1.28_{-0.29}^{+0.27}+0.04 i$\\
$a_{K \Xi}^{(1)}  $&$=$&$+0.15$&$+0.34_{-0.18}^{+0.17}$&$+0.22_{-0.03}^{+0.02}+0.40 i$&$+0.72_{-0.21}^{+0.19}+0.40 i$\\
$a_{K \Xi}^{(0)}  $&$=$&$+0.66$&$+0.54_{-0.22}^{+0.18}$&$+0.29_{-0.10}^{+0.06}+0.24 i$&$+1.48_{-0.32}^{+0.24}+0.24 i$\\
$a_{\bar{K} \Xi}^{(1)} $&$=$&$-0.50$&$+0.44_{-0.07}^{+0.04}$&$-0.27_{-0.04}^{+0.07}$&$-0.33_{-0.11}^{+0.11}$\\
$a_{\bar{K} \Xi}^{(0)} $&$=$&$-0.15$&$+0.24_{-0.29}^{+0.30}$&$+0.38_{+0.00}^{+0.00}$&$+0.48_{-0.29}^{+0.30}$\\
$a_{K \Lambda}^{(1/2)}$&$=$&$-0.04$&$+0.48_{-0.19}^{+0.18}$&$-0.10_{+0.00}^{+0.00}+0.34 i$&$+0.34_{-0.19}^{+0.18}+0.34 i$\\
$a_{\bar{K} \Lambda}^{(1/2)}$&$=$&$-0.05$&$+0.48_{-0.19}^{+0.18}$&$-0.10_{+0.00}^{+0.00}+0.34 i$&$+0.32_{-0.19}^{+0.18}+0.34 i$\\
\hline
$a_{\eta N}^{(1/2)}$&$=$&$-0.01$&$+0.24_{-0.25}^{+0.23}$&$+0.08_{+0.00}^{+0.00}+0.30 i$&$+0.31_{-0.25}^{+0.23}+0.30 i$\\
$a_{\eta \Xi}^{(1/2)}$&$=$&$-0.09$&$+0.83_{-0.27}^{+0.26}$&$-0.01_{+0.00}^{+0.00}+0.34 i$&$+0.73_{-0.27}^{+0.26}+0.34 i$\\
$a_{\eta \Sigma}^{(1)}$&$=$&$-0.04$&$+0.23_{-0.08}^{+0.08}$&$+0.06_{+0.00}^{+0.00}+0.22 i$&$+0.25_{-0.08}^{+0.08}+0.22 i$\\
$a_{\eta \Lambda}^{(0)}$&$=$&$-0.04$&$+0.29_{-0.19}^{+0.16}$&$+0.07_{+0.00}^{+0.00}+0.64 i$&$+0.32_{-0.19}^{+0.16}+0.64 i$\\
\hline
\hline
\end{tabular}
\renewcommand{\arraystretch}{1.0}
\caption{\textit{Truncated} results for the fit to $a^{+}_{\pi N}$, $a_{K
    N}^{(1)}$, $a_{K N}^{(0)}$, and $d_0=-0.996$~GeV$^{-1}$. All scattering
  lengths are given in units of fm.  $\Sigma$ denotes  the sum of all contributions from first, second, and
  third order.}\label{tab:truncated-fit}
\end{center}
\end{table*}

\subsubsection{Reordering procedure}

We also have employed different sets of parameters in~\eqref{eqn:mapping}, i.e.\ 
obtained from slightly different fitting procedures, and used results from
resonance saturation.  None of these show better agreement with the
data and thus we refrain from giving detailed results for these alternative fits.

Another method to improve the convergence behavior, called reordering of the 
chiral series, was proposed in Ref.~\cite{Becher:1999he} 
and applied to SU(3) ChPT in Ref.~\cite{Mojzis:1999qw}. 
There the seemingly slow convergence of the chiral series for e.g the
$\Lambda$-mass is discussed and traced back to the question whether the
direct comparison of the different orders of the chiral series for some
observable is the best choice.  Instead it might be reasonable to express the 
results not in bare quantities but in terms of physical observables.
From the practical point of view this means that the low-energy constants 
are at least symbolically expanded in a chiral series. 
As before we split the fitting procedure into two steps. 
We rewrite the baryon mass in the chiral limit and the symmetry breaking LECs as
\begin{align}
m_0 &=m^{(0)}+m^{(1)}+m^{(2)}+\ldots \,,\nn
b_0 &=b_0^{(0)}+b_0^{(1)}+\ldots \,, \nn
b_D &=b_D^{(0)}+b_D^{(1)}+\ldots \,, \nn
b_F &=b_F^{(0)}+b_F^{(1)}+\ldots \,.
\end{align}
First of all we  fix $m^{(0)}=1.150$~GeV, then $m^{(1)}$, $b_0^{(0)}$,
$b_D^{(0)}$ and $b_F^{(0)}$ are fitted to the physical values of the baryon masses. 
Finally using these constants we fit the third-order mass corrections
including $m^{\IR/\HB,(2)}$, $b_0^{\IR/\HB,(1)}$, $b_D^{\IR/\HB,(1)}$ and $b_F^{\IR/\HB,(1)}$ again to the
baryon masses and the $\pi N$ sigma term in the IR and HB regime, respectively. For the central values we obtain
\begin{align}
m^{(1)} &=+0.02\text{ GeV} ~,		 &m^{\IR,(2)}  &=+0.22\text{ GeV}~,	\nn &&m^{\HB,(2)} &=+0.25\text{ GeV}~,	\nn
b_0^{(0)}&=-0.02\text{ GeV}^{-1} ~,	 &b_0^{\IR,(1)}&=-0.22\text{ GeV}^{-1}~,	\nn &&b_0^{\HB,(1)}&=-0.25\text{ GeV}^{-1}~,	\nn
b_D^{(0)}&=+0.07\text{ GeV}^{-1} ~,	 &b_D^{\IR,(1)}&=-0.02\text{ GeV}^{-1}~,	\nn &&b_D^{\HB,(1)}&=+0.11\text{ GeV}^{-1}~,	\nn
b_F^{(0)}&=-0.20\text{ GeV}^{-1}~,	 &b_F^{\IR,(1)}&=-0.26\text{ GeV}^{-1}~,	\nn &&b_F^{\HB,(1)}&=-0.66\text{ GeV}^{-1}~.	
\end{align}
In the same manner, the dynamical LECs are subjected to a chiral expansion
\beq
B_i = B_i^{(0)}+B_i^{(1)}+\ldots\quad (i = 1,2,3,4) ~.
\eeq
We fix the constants $\{B_1^{(0)},\ldots,B_4^{(0)}\}$ by fitting the results up 
to second chiral order to the experimental results for $a^{+}_{\pi N}$, $a_{K N}^{(1)}$, $a_{K N}^{(0)}$,  
and $d_0=-0.996$~GeV$^{-1}$. 
The results read
\begin{align}
B_1^{(0)} &=(+0.35\pm 0.00)\text{ GeV}^{-1}~, \nn B_2^{(0)} &=(-0.20\pm 0.09)\text{ GeV}^{-1}~, \nn
B_3^{(0)} &=(-0.11\pm 0.09)\text{ GeV}^{-1}~, \nn B_4^{(0)} &=(-0.04\pm 0.09)\text{ GeV}^{-1}~.
\end{align}
Subsequently, using these LECs, we fix $\{B_1^{(1)},\ldots,B_4^{(1)}\}$ by fitting the
\begin{table*}
\begin{center}
\renewcommand{\arraystretch}{1.5}
\begin{tabular}{rcclll}
\hline
\hline
Channel&$=$&$\Order(q^1) $&$+\Order(q^2)$&$+\Order(q^3)_{\IR}$&$\qquad\sum_{\IR}$\\
\hline
$a_{K N}^{(1)}  $&$=$&$\qquad-0.45\qquad$&$+0.12_{+0.00}^{+0.00}\qquad$&$+0.00_{-0.30}^{+0.30}\qquad$&$-0.33_{-0.30}^{+0.30}$\\
$a_{K N}^{(0)}  $&$=$&$+0.04$&$-0.02_{-0.14}^{+0.14}$&$+0.00_{-0.76}^{+0.76}$&$+0.02_{-0.90}^{+0.90}$\\
$a_{\bar{K} N}^{(1)} $&$=$&$+0.20$&$+0.05_{-0.07}^{+0.07}$&$-0.09_{-0.49}^{+0.43}+0.18 i\qquad$&$+0.16_{-0.56}^{+0.50}+0.18 i$\\
$a_{\bar{K} N}^{(0)} $&$=$&$+0.53$&$+0.19_{-0.07}^{+0.07}$&$+0.39_{-0.63}^{+0.52}+0.22 i$&$+1.11_{-0.70}^{+0.59}+0.22 i$\\
\hline
\hline
\end{tabular}
\renewcommand{\arraystretch}{1.0}
\caption{\textit{Full} result for the reordered fit to $a^{+}_{\pi N}$, $a_{K N}^{(1)}$, $a_{K N}^{(0)}$, 
and $d_0=-0.996$GeV$^{-1}$. All scattering lengths are given in units of 
fm. $\Sigma$ denotes  the sum of all contributions from first, second, and
  third order.  }\label{tab:full-reord}
\end{center}
\end{table*}
results for the scattering lengths up to  third order, including the
\textit{truncated} and the \textit{full} loop corrections, to the same 
experimental results.  Thus at this order we obtain two sets of constants 
for the heavy-baryon and the covariant approach, respectively,
as follows:
\begin{align}
B_1^{\IR,(1)}&=+0.49_{-0.12}^{+0.08}\text{ GeV}^{-1}~, \nn
B_2^{\IR,(1)}&=-0.56_{-0.33}^{+0.38}\text{ GeV}^{-1}~, \nn
B_3^{\IR,(1)}&=+0.31_{-0.49}^{+0.42}\text{ GeV}^{-1}~, \nn
B_4^{\IR,(1)}&=-0.32_{-0.53}^{+0.56}\text{ GeV}^{-1}~, \nn
B_1^{\HB,(1)}&=+0.31_{-0.04}^{+0.02}\text{ GeV}^{-1} ~,\nn
B_2^{\HB,(1)}&=-0.16_{-0.22}^{+0.23}\text{ GeV}^{-1} ~,\nn
B_3^{\HB,(1)}&=-0.15_{-0.27}^{+0.25}\text{ GeV}^{-1} ~,\nn
B_4^{\HB,(1)}&=-0.04_{-0.28}^{+0.29}\text{ GeV}^{-1} ~.
\end{align}
Using these reordered low-energy constants we obtain the results for the 
(anti)kaon--nucleon scattering lengths as presented in
Tables~\ref{tab:full-reord} and \ref{tab:truncated-reord}. 
However no clear improvement of the convergence behavior is achieved by 
reordering the chiral series (this statements also holds for the other
channels not given in the tables). On the other hand using the formulas 
from Ref.~\cite{Bernard:1993nj} the convergence rate of the chiral series for the 
baryon masses is tremendously improved~\cite{Mojzis:1999qw}, 
which we also checked.  Also the convergence of the chiral series for the 
baryon magnetic moments and axial couplings was shown there to be 
improved by the reordering procedure. The different response of the chiral 
series for the scattering lengths is presumably related to the fact that
we did not consider the local  dimension-three operators with adjustable
LECs here. This requires further investigation.

\begin{table*}
\begin{center}
\renewcommand{\arraystretch}{1.5}
\begin{tabular}{rcclll}
\hline
\hline
Channel&$=$&$\Order(q^1) $&$+\Order(q^2)$&$+\Order(q^3)_{\HB}$&$\qquad\sum_{\HB}$\\
\hline
$a_{K N}^{(1)}  $&$=$&$\qquad-0.45\qquad$&$+0.12_{+0.00}^{+0.00}\qquad$&$+0.00_{-0.10}^{+0.10}\qquad$&$-0.33_{-0.10}^{+0.10}$\\
$a_{K N}^{(0)}  $&$=$&$+0.04$&$-0.02_{-0.14}^{+0.14}$&$+0.00_{-0.40}^{+0.40}$&$+0.02_{-0.54}^{+0.54}$\\
$a_{\bar{K} N}^{(1)} $&$=$&$+0.20$&$+0.05_{-0.07}^{+0.07}$&$+0.15_{-0.26}^{+0.24}+0.36 i\qquad$&$+0.40_{-0.33}^{+0.31}+0.36 i$\\
$a_{\bar{K} N}^{(0)} $&$=$&$+0.53$&$+0.19_{-0.07}^{+0.07}$&$+0.80_{-0.35}^{+0.29}+0.22 i$&$+1.52_{-0.42}^{+0.36}+0.22 i$\\
\hline
\hline
\end{tabular}
\renewcommand{\arraystretch}{1.0}
\caption{\textit{Truncated} results for the reordered fit to 
$a^{+}_{\pi N}$, $a_{K N}^{(1)}$, $a_{K N}^{(0)}$, and
$d_0=-0.996$~GeV$^{-1}$. All scattering lengths are given in units of fm.
$\Sigma$ denotes  the sum of all contributions from first, second, and
  third order.}\label{tab:truncated-reord}
\end{center}
\end{table*}

\subsubsection{Discussion}

The conclusion to be drawn based on the numerical evaluation above is that 
the chiral series of the scattering lengths does not appear to converge in general. The
best convergence rate is observed  for the pion channels in the truncated approach. 
The reason for the slower convergence for SU(3) compared to  SU(2) is clearly 
the magnitude of the expansion parameter, being  $M_K/m_0\sim M_\eta/m_0\sim
1/2$ and $M_\pi/m_0\sim 1/7$ for the  SU(3) and SU(2) case, respectively.
Moreover for the truncated (HBChPT) result,  the convergence of the chiral
series is improved at least for the pion channels in comparison to the full results. 
A similar observation was made  in Ref.~\cite{Torikoshi:2002bt} in the analysis of
pion--nucleon scattering in both schemes.  On the other hand
in Ref.~\cite{Ellis:1999jt} the loop corrections for the baryon masses within SU(3) ChPT 
seem to be smaller in the infrared than in the HB approach.  Of course, for a
definitive statement on the convergence of the chiral expansion one needs to
include the so-far neglected finite pieces of the local dimension-three operators.
These certainly are expected to lead to cancellations at third order, reducing
the overall size of these corrections. This needs to be checked explicitly in
future calculations. Our purpose here, however, was to simply compare the
results of the covariant and the heavy-baryon approach under the same set of
assumptions. To further make progress in SU(3) meson--baryon scattering,
unitary coupled-channel calculations  properly matched to the ChPT amplitudes
derived here should be performed. Furthermore, the explicit representation of
the scattering amplitudes in SU(2) and SU(3) allows us to address the issue of
matching, which we now turn to and consider the central part of this work.

\section{Matching to SU(2)}
\label{sec:match}

A vast body of work has been performed on matching the two different chiral expansions 
in the meson sector, beginning with leading-order matching
of the $\Order(q^4)$ low-energy constants~\cite{GL:NPB250}.
Several constants have by now been matched at two-loop 
level~\cite{Moussallam,KaiserSchweizer,Schmid:Thesis,match:p4,match:p6},
and there are results for the electromagnetic~\cite{pipiatom,Sazdjian,match:EM}
and the anomalous~\cite{KampfMouss} sector.  
In all cases, the relations found can be used to transfer information on the coupling
constants from the two-flavor theory to the three-flavor one and vice versa.
Typically, more precise phenomenological information on the SU(2) sector of the theory
can be used to constrain certain linear combinations of SU(3) low-energy constants, 
while model estimates (like resonance saturation) are often performed in SU(3) 
and have to be translated in the opposite direction in order to be used in SU(2) calculations.

Remarkably little is known about such matching relations in ChPT with baryons,
beyond the trivial leading-order relations as the well-known one for 
the axial coupling constants $g_A = D+F+\Order(M_K^2)$.
A major study has been performed on the baryon masses up to $\Order(q^4)$~\cite{Frink:2004ic}.
Here we complete the matching for all constants of the $\Order(q^2)$ pion--nucleon Lagrangian
to the first nontrivial order, including $\Order(M_K)$ corrections.

\subsection{Low-energy constants in the pion--nucleon sector}

Pion--nucleon scattering can be analyzed in the framework of SU(2) as well as SU(3) ChPT. 
The second-order SU(2) ChPT pion--nucleon Lagrangian reads
\begin{align}
\mathcal{L}^{(1+2)}_{\pi N}&=\bar N\Big\{i \slashed{D}-\krig{m}_N +\frac{g}{2}\slashed{u}\gamma_5
+c_1\langle\chi_+\rangle\nn
&\qquad-\frac{c_2}{8\krig{m}_N^2}\bigl(\langle u_\mu u_\nu\rangle \{D^\mu,D^\nu\}+
{\rm h.c.}\bigr)\nn
&\qquad+\frac{c_3}{2}\langle u_\mu u^\mu\rangle
+\frac{c_4}{4} [u_\mu,u_\nu]i\sigma^{\mu \nu}
+c_5\tilde\chi_+\nn
&\qquad+ \frac{1}{8\krig{m}_N} \bigl( c_6 \F_{\mu\nu}^+ 
+ c_7 \langle \F_{\mu\nu}^+ \rangle \bigr) \sigma^{\mu\nu}\Big\}N ~. \label{eq:LpiN}
\end{align}
$\chi = 2B\,\textrm{diag}(m_u,m_d) + \ldots$ now contains the light quark mass
matrix only, $B$ is related to the light (up, down) quark condensate in the chiral limit,
and $\tilde \Order = \Order - \langle\Order\rangle/2$ refers to the traceless 
part of the operator $\Order$.
$\krig{m}_N$ and $g$ are the nucleon mass and axial coupling constant in the chiral limit, respectively.
For an external electromagnetic field $A_\mu$, the field strength tensor reads
$\F_{\mu\nu} = (\partial_\mu A_\nu-\partial_\nu A_\mu) \mathcal{Q}$ 
with the \emph{nucleon} charge matrix $\mathcal{Q}=e\,\mathrm{diag}(1,0)$,
and $\F^+_{\mu\nu} = u^\dagger \F_{\mu\nu} u + u \F_{\mu\nu} u^\dagger$.
Out of the second-order LECs in~\eqref{eq:LpiN}, only $c_{1-4}$ feature in $\pi N$ scattering
in the isospin limit, so we  include $c_{5-7}$ and the matching relations for those 
obtained from other sources for completeness.

The very meaning of the low-energy constants in effective field theories goes 
back to the fact that  the heavy degrees of freedom are integrated out. 
In this way, the strange quark can be integrated out such that three-flavor ChPT reduces to the two-flavor theory. 
This means that to obtain the next-to-leading-order matching relations for the LECs
we calculate the scattering amplitude for pion--nucleon scattering to one loop, 
where we only take SU(3)/SU(2) intermediate states into account, 
which can be represented symbolically as
\begin{align}
\sum{c_i}=\sum{b_i}+\F[\text{SU(3)/SU(2)-states}]+\Order(M_K^2) ~,\label{eqn:motiv}
\end{align}
where the loop functional $\F$ contains contributions from the loop 
diagrams in Fig.~\ref{pic:graphs} as well as wave-function renormalization.

To assign the results to the various structures in~\eqref{eq:LpiN} 
one realizes first of all that the LECs can be distinguished by their associated Dirac structures, 
i.e.\ $c_4$ ($b_5,b_6,b_7$) appears in association with $\sigma^{\mu \nu}$, 
in contrast to the remaining LECs escorting trivial Dirac structures. 
The latter can be identified as the coefficients of the second-order energy structures 
$(t-2M_\pi^2)$, $M_\pi^2$ and $\XX^2:=(s-m_0^2)^2$. 
To assign the one-loop contributions to the LECs correctly we first modify  
\eqref{eqn:T} since $T_1^{b,j,i,a}$ and $T_2^{b,j,i,a}$ do not individually fulfill the power counting.
This did not matter in Sec.~\ref{sec:MBscatt} since the lowest-order contributions 
cancelled for the scattering lengths. Here we use
\begin{align}
T^{b,j,i,a}&=T_3^{b,j,i,a}+[\slashed{q}_j,\slashed{q}_i] T_4^{b,j,i,a} ~, \label{eqn:TT}\\
T_3&=T_1+\Big(m_0+\frac{s-u}{4m_0}\Big)T_2 ~, \quad 
T_4=-\frac{1}{4m_0}T_2 ~.\nonumber
\end{align}
Now for a more delicate matter. 
The loop integrals must be evaluated properly, i.e.\ taking into account the
different expansion parameters in both theories. 
We start with the loop integrals as defined in IR-regularized SU(3) ChPT. 
Having performed the loop integration,  we expand the integrand of the remaining parameter integral, 
see also Appendix~\ref{App:thr-exp}, to a certain order in $(t-2M_\pi^2)$, $M_\pi^2$, and $\XX^2$. 
Afterwards we expand the final result in $M_{K,\eta}$, which is necessary
because the one-loop result only fully determines 
the first-order corrections to the matching relations. 
This procedure is called double-scale expansion, where we assume 
$m_0\gg M_{K,\eta}\gg M_\pi$. 
The results for the expanded integrals are presented in Appendix~\ref{App:MatIntr}.

Here we present the next-to-leading-order constraints on the LECs of SU(2) ChPT $c_{1-7}$.
The expression for $c_5$ is given in Ref.~\cite{Frink:2004ic},
those for $c_{6/7}$ were derived from the SU(3) analysis of baryon form factors in Ref.~\cite{Kubis:BaryonFF}, 
but ultimately go back to Ref.~\cite{CaldiPagels}.
The relation for $c_1$ can be uniquely determined from meson--baryon scattering -- 
the result is fully consistent with the analysis of the baryon masses~\cite{Frink:2004ic}.
The results for $c_{2-4}$ are new and have not been given before.
For completeness, we also include the matching relations for the leading-order constants $\krig{m}_N$ and $g$.
We recall that due to the Gell-Mann--Okubo relation
\begin{align}
M^2_{\eta}=\frac{4}{3}M_K^2+\Order(M^2_\pi)~,\label{eqn:GMO}
\end{align}
only one heavy meson mass is needed.  Moreover we use~\cite{GL:NPB250}
\begin{equation}
l_4^r=8L_4^r+4L_5^r-\frac{1}{64\pi^2}\biggl(2\log\frac{M_K}{\mu}+1\biggr) ~.
\end{equation}
Finally, in order to complete the matching for the magnetic moments LECs
$c_{6/7}$, we have to amend the Lagrangian $\Lagr^{(2)}_{\phi B}$~\eqref{eq:L2} with 
the corresponding SU(3) structures,
\beq
\Lagr^{(2)}_{\phi B} = \ldots +  b_{12/13} \bigl\langle \bar{B} 
\sigma^{\mu\nu} [F_{\mu\nu}^+,B]_\mp \bigr\rangle ~,
\eeq
where
$F^+_{\mu\nu} = u^\dagger F_{\mu\nu} u + u F_{\mu\nu} u^\dagger$, 
$F_{\mu\nu} = (\partial_\mu A_\nu-\partial_\nu A_\mu) Q$ with the quark charge matrix
$Q=e\,\mathrm{diag}(2,-1,-1)/3$.
Altogether the matching relations read
\begin{widetext}
\begin{align}
\krig{m}_N &= m_0 -4M_K^2 \bigl(b_0+b_D-b_F\bigr) 
- \frac{M_K^3}{48\pi \Fpi^2}\biggl[5D^2-6DF+9F^2+\frac{4}{3\sqrt{3}}(D-3F)^2\biggr]
+ \Order(M_K^4) ~, \nn
g &= D+F + \Order(M_K^2) ~, \quad 
c_1=b_0+\frac{b_D}{2}+\frac{b_F}{2}  
+ \frac{M_K}{256\pi \Fpi^2}\bigg[5D^2-6DF+9F^2+\frac{2}{3\sqrt{3}}(D-3F)^2\bigg]
+ \Order(M_K^2) ~,\nn
c_2&=b_8+b_9+b_{10}+2b_{11}
- \frac{M_K}{128 \pi F_\pi^2}\bigg[6+\frac{19}{3}D^4+4 D^3F+\frac{58}{3} D^2 F^2-12 D F^3+25 F^4 
-\frac{8 (D-3 F)^2 (D+F)^2}{3\sqrt{3}}\bigg] \nn &\quad
+\Order(M_K^2) ~, \nn
c_3&=b_1+b_2+b_3+2b_4 
+\frac{M_K}{128 \pi F_\pi^2}\bigg[5D^2 -6DF+9F^2+ \frac{19}{3} D^4+4 D^3F+\frac{58}{3} D^2 F^2-12 D F^3 +25 F^4 
\nn &\quad
+\frac{8 (D-3 F)^2 (D+F)^2}{3\sqrt{3}}\bigg] +\Order(M_K^2) ~,\nn
c_4&=4(b_5+b_6)
+\frac{M_K}{96\pi \Fpi^2}\bigg[D^2-6DF-3F^2
-\frac{9}{2} D^4-10 D^3F+ D^2 F^2-18 DF^3 -\frac{33}{2} F^4
-\frac{2(D-3F)^2(D+F)^2}{\sqrt{3}} \bigg] \nn &\quad
+\Order(M_K^2)~,\nn
c_5&= b_D+b_F
-\frac{M_K}{128\pi \Fpi^2}\bigg[D^2-6DF-3F^2+\frac{8}{3\sqrt{3}}(D-3F)(D+F)\bigg]
+\Order(M_K^2)~,\nn
\frac{c_6}{\krig{m}_N}&=8\big(b_{12}+b_{13}\big) + \frac{M_K}{24\pi \Fpi^2}\big[D^2-6DF-3F^2\big]
+\Order(M_K^2)~, \quad 
\frac{c_7}{\krig{m}_N} =-\frac{16}{3}b_{13} -\frac{ M_K}{8\pi \Fpi^2}(D-F)^2
+\Order(M_K^2)~. \label{eq:ciMatching}
\end{align}
\end{widetext}

The shifts $\Delta c_i$ of $\Order(M_K)$ in the matching relations~\eqref{eq:ciMatching} 
are finite and calculable in terms of well-known parameters.
In particular, for the constants relevant in $\pi N$ scattering, we find 
$\Delta c_1 = +0.2$\,GeV$^{-1}$, $\Delta c_2 = -2.1$\,GeV$^{-1}$, 
$\Delta c_3 = +1.6$\,GeV$^{-1}$, $\Delta c_4 = +2.0$\,GeV$^{-1}$, 
hence these shifts are rather sizeable.

\subsection{Low-energy constants in the pion--hyperon sector}

Chiral SU(2) has been used to describe hyperon properties in Refs.~\cite{Tiburzi:2008bk,Jiang:2009sf}, 
focusing on the chiral expansion of the masses and axial coupling constants.  
In the same spirit as the SU(2) description of pion--kaon systems~\cite{Roessl,Frink:piK},
one expects the convergence properties of the theory to be improved; on the other hand,
as one is restricted to strangeness-conserving processes, there are typically less observables
related by chiral symmetry, or more low-energy constants to be fixed.

In the following, we construct the \emph{complete} next-to-leading order Lagrangians
for the $\pi\Sigma$, $\pi\Lambda$, and $\pi\Xi$ systems. 
In analogy to the $\pi N$ system, we denote the second-order LECs by
$c_i^\Sigma$, $c_i^\Lambda$, and $c_i^\Xi$, respectively.
The leading-order matching relations on these constants are already determined by the Lagrangians themselves. 
The next-to-leading-order is fully determined by the analysis of the one-loop corrections in the SU(3) and 
SU(2) theories in the sense of~\eqref{eqn:motiv}, where we refer to the corresponding subgroup SU(2) of the SU(3).
For the results of the expanded loop integrals, see Appendices~\ref{App:Integrals-results} and \ref{App:MatIntr}.

We start out with the strangeness $S=1$ sector of the theory.
Although the $\Sigma$ and $\Lambda$ hyperons belong to different 
isospin multiplets, strangeness-neutral currents induce transitions between
the two sectors, such that the Lagrangian is of the form
\beq
\Lagr_{S=1} = \Lagr_{\pi \Lambda} + \Lagr_{\pi \Sigma} + \Lagr_{\pi \Lambda\Sigma} ~. 
\eeq
We find
\begin{align}
\Lagr_{\pi \Lambda}^{(1+2)} &= \bar \Lambda \biggl\{ i \slashed{\partial}- \krig{m}_\Lambda 
+c_1^\Lambda \langle \chi_+ \rangle
\nn&\qquad-\frac{c_2^\Lambda}{8\krig{m}_\Lambda^2} \bigl( \langle u_\mu u_\nu \rangle \{\partial^\mu,\partial^\nu\} + {\rm h.c.} \bigr)
\nn&\qquad+\frac{c_3^\Lambda}{2} \langle u_\mu u^\mu \rangle 
+\frac{c_7^\Lambda}{8\krig{m}_\Lambda}  \langle F_{\mu\nu}^+ \rangle \sigma^{\mu\nu} \biggr\} \Lambda ~, \label{eq:LpiLambda} \\
\Lagr_{\pi\Sigma}^{(1+2)}&=\langle \bar\Sigma \bigl(i\slashed{D}- \krig{m}_\Sigma \bigr)\Sigma \rangle
+ \frac{g_\Sigma}{2} \langle \bar\Sigma \gamma^\mu\gamma_5[u_\mu,\Sigma] \rangle
\nn&
+c_1^\Sigma \langle \chi_+ \rangle \langle \bar\Sigma \Sigma \rangle
\nn&
-\frac{c_{2a}^\Sigma}{8\krig{m}_\Sigma^2} \bigl( \langle u_\mu u_\nu \rangle \langle \bar\Sigma \{D^\mu,D^\nu\}\Sigma \rangle + {\rm h.c.} \bigr)
\nn&
-\frac{c_{2b}^\Sigma}{8\krig{m}_\Sigma^2} \bigl( \langle \bar\Sigma u_\nu \rangle \langle u_\mu \{D^\mu,D^\nu\}\Sigma \rangle + {\rm h.c.} \bigr)
\nn&+\frac{c_{3a}^\Sigma}{2} \langle \bar\Sigma \Sigma \rangle \langle u_\mu u^\mu \rangle
+\frac{c_{3b}^\Sigma}{2} \langle \bar\Sigma u_\mu \rangle \langle u^\mu \Sigma \rangle\nn
&
+ \frac{c_4^\Sigma}{4} \langle \bar\Sigma u_\mu\rangle \langle u_\nu \Sigma \rangle i\sigma^{\mu\nu}
+ c_5^\Sigma \langle \bar\Sigma [\tilde\chi_+,\Sigma]\rangle \label{eq:LpiSigma} \\
&+ \frac{c_6^\Sigma}{8\krig{m}_\Sigma} \langle\bar\Sigma \sigma^{\mu\nu}[\tilde F_{\mu\nu}^+,\Sigma]\rangle 
+ \frac{c_7^\Sigma}{8\krig{m}_\Sigma} \langle\bar\Sigma \sigma^{\mu\nu} \Sigma \rangle \langle F_{\mu\nu}^+\rangle ~, \nn
\Lagr_{\pi\Lambda\Sigma}^{(1+2)} &=  
\Bigl(\frac{g_{\Lambda\Sigma}}{2} \langle \bar\Sigma u_\mu\rangle \gamma^\mu\gamma_5 \Lambda +{\rm h.c.} \Bigr)
\nn&+ \frac{c_6^{\Lambda\Sigma}}{4(\krig{m}_\Lambda+\krig{m}_\Sigma)} \bigl(\langle\bar\Sigma F_{\mu\nu}^+ \rangle \sigma^{\mu\nu} \Lambda + {\rm h.c.}\bigr) ~.
\end{align}
We use the field strength tensor $F_{\mu\nu} = (\partial_\mu A_\nu-\partial_\nu A_\mu) Q$, 
where $Q=e\,\mathrm{diag}(2,-1)/3$ is the SU(2) \emph{quark} charge matrix. 
As the $\Lambda$ is an isoscalar particle, $\Lagr_{\pi\Lambda}^{(1+2)}$ contains less terms than 
the corresponding pion--nucleon Lagrangian;
note, in particular, the absence of a covariant derivative and of a axial-vector-type term. 
As the $I=1$ triplet of $\Sigma$ fields,
\beq
\Sigma = \frac{1}{\sqrt{2}}\begin{pmatrix}
\Sigma^0  & \sqrt{2}\Sigma^+  \\
\sqrt{2}\Sigma^- & - \Sigma^0
\end{pmatrix} ~, 
\eeq 
transforms in the adjoint representation, \eqref{eq:LpiSigma} has been constructed 
starting from the SU(3) Lagrangian, using additional matrix trace relations for SU(2). 
We find that only the $c_2$ and $c_3$ type terms are ``doubled'' compared to
the  pion--nucleon Lagrangian,  otherwise the number of terms stays the same. 
Finally for $\Lagr_{\pi\Lambda\Sigma}^{(1+2)}$, 
we assume the mass term to be diagonalized for the physical $\Lambda$ and
$\Sigma^0$ fields already, hence we do not include a term 
$\propto\langle\bar\Sigma\tilde\chi_+\rangle\Lambda$.  

We briefly comment on the different (chiral limit) masses $\krig{m}_\Sigma$, $\krig{m}_\Lambda$ 
in the Lagrangians~\eqref{eq:LpiLambda}, \eqref{eq:LpiSigma}.
In the framework of chiral SU(3), one finds $\krig{m}_\Sigma - \krig{m}_\Lambda = \Order(M_K^2)$
(see the matching formulas below), so the mass
difference does not vanish in the SU(2) chiral limit, and formally one would have to treat
 $\krig{m}_\Sigma - \krig{m}_\Lambda$ as a large quantity in chiral SU(2).  However for physical values of the 
quark masses, this mass difference is smaller than the pion mass, and the authors of Ref.~\cite{Tiburzi:2008bk}
treat it, in a phenomenological counting scheme, as $\Order(M_\pi)$
(comparable to the inclusion of the decuplet, where one often counts the decuplet--octet
mass difference as $\Order(M_\pi)$), a procedure we will also adopt in the following.
This counting scheme may become problematic when considering a regime with
$M_\pi \ll m_\Sigma-m_\Lambda \approx 87$~MeV, but for physical (and larger) pion masses,
it presents a very useful approach.

The matching relations for the terms in $\Lagr_{\pi \Lambda}^{(1+2)}$
including terms  of $\Order(M_K)$  read
\begin{widetext}
\begin{align}
\krig{m}_\Lambda &= m_0 -4M_K^2 \Bigl(b_0+\frac{4}{3}b_D\Bigr) 
-\frac{M_K^3}{24\pi \Fpi^2}\biggl[\Bigl(1+\frac{8}{3\sqrt{3}}\Bigr)D^2+9F^2\biggr]
+ \Order(M_K^4) ~, \nn
c_1^\Lambda&=b_0+\frac{b_D}{3} 
+\frac{M_K}{128\pi F_\pi^2}\bigg[\Bigl(1+\frac{4}{3\sqrt{3}}\Bigr)D^2+9F^2\bigg]
+\Order(M_K^2) ~,\nn
c_2^\Lambda&=\frac{4}{3}b_{10}+2b_{11}
-\frac{3M_K}{32\pi\Fpi^2}\bigg[1+
\Big(\frac{11}{2}+\frac{16}{3\sqrt{3}}\Big) \frac{D^4}{9}+\frac{41}{9} D^2 F^2+\frac{3}{2} F^4
\bigg] +\Order(M_K^2) ~,\nn
c_3^\Lambda&=\frac{4}{3}b_3+2b_4
+\frac{3M_K}{32\pi\Fpi^2}\bigg[\frac{D^2}{6}+\frac{3}{2}F^2+
\Big(\frac{11}{2}+\frac{16}{3\sqrt{3}}\Big) \frac{D^4}{9}+\frac{41}{9} D^2 F^2+\frac{3}{2} F^4
\bigg] +\Order(M_K^2) ~,\nn
\frac{c_7^\Lambda}{\krig{m}_\Lambda}&=-8b_{13} +\frac{3M_K}{4\pi \Fpi^2}DF+\Order(M_K^2) ~. \label{eq:matchL}
\end{align}
For the next-to-leading-order constraints on the low-energy constants in the $\pi\Sigma$ Lagrangian we find
\begin{align}
\krig{m}_\Sigma &= m_0 -4M_K^2 b_0  
-\frac{M_K^3}{8\pi \Fpi^2}\biggl[\Bigl(1+\frac{8}{3\sqrt{3}}\Bigr)D^2+F^2\biggr]
+ \Order(M_K^4) ~, \quad
g_\Sigma =	 2F + \Order(M_K^2) ~, \nn
c_1^\Sigma&=b_0+b_D
+\frac{3M_K}{128\pi F_\pi^2}\bigg[\Bigl(1+\frac{4}{9\sqrt{3}}\Bigr)D^2+F^2\bigg]
+\Order(M_K^2) ~,\nn
c_{2a}^\Sigma&=b_8+\frac{b_{11}}{2}
-\frac{M_K}{128\pi\Fpi^2}\bigg[
1+\frac{3}{2} D^4 + \Big(1+\frac{32}{3\sqrt{3}}\Big)D^2 F^2 +\frac{19}{2} F^4
\bigg] +\Order(M_K^2) ~,\nn
c_{2b}^\Sigma&=\frac{1}{2}(b_{10}-b_{8})
-\frac{M_K}{96 \pi F_\pi^2}\bigg[
(D^2-3F^2) F^2 +\frac{2D^2(D^2-6 F^2)}{3\sqrt{3}}\bigg]
+\Order(M_K^2) ~,\nn
c_{3a}^\Sigma&=2b_1+b_4
+\frac{3M_K}{128\pi\Fpi^2}\bigg[D^2+F^2 + D^4 +\frac{2}{3}\Big(1+\frac{32}{3\sqrt{3}}\Big)D^2 F^2
+ \frac{19}{3} F^4 
\bigg]
+\Order(M_K^2) ~,\nn
c_{3b}^\Sigma&=2b_3-b_1
+\frac{M_K}{24  \pi F_\pi^2}\bigg[(D^2-3 F^2)F^2+\frac{2 D^2 (D^2-6 F^2)}{3 \sqrt{3}}\bigg]
+\Order(M_K^2) ~,\nn
c_4^\Sigma&= 4b_{5}+b_{7}
-\frac{M_K}{32\pi\Fpi^2}\bigg[D^2+F^2 -
\frac{D^4}{2} + \Big( 19 + \frac{8}{\sqrt{3}}\Big)\frac{D^2 F^2}{3} +\frac{7}{2} F^4 
\bigg] +\Order(M_K^2) ~,\nn
c_5^\Sigma &= b_F  + \frac{3M_K}{32\pi \Fpi^2}\Bigl(1+\frac{8}{9\sqrt{3}}\Bigr)DF
+\Order(M_K^2) ~,\nn
\frac{c_6^\Sigma}{\krig{m}_\Sigma} &= 8b_{12} - \frac{M_K}{8\pi \Fpi^2}(D^2+F^2) 
+\Order(M_K^2) ~, \quad 
\frac{c_7^\Sigma}{\krig{m}_\Sigma} = 8b_{13} - \frac{3M_K}{4\pi \Fpi^2}DF 
+\Order(M_K^2) ~. \label{eq:matchS}
\end{align}
Finally, the matching of the transition couplings does not contain new information
from meson--baryon scattering, we simply have
\beq
g_{\Lambda\Sigma} = \frac{2}{\sqrt{3}} D + \Order(M_K^2) ~, \quad
\frac{c_6^{\Lambda\Sigma}}{\krig{m}_\Lambda+\krig{m}_\Sigma} = \frac{4}{\sqrt{3}} b_{13} 
- \frac{DF}{8\sqrt{3}\pi \Fpi^2}M_K + \Order(M_K^2) ~. \label{eq:matchLS}
\eeq
The relations for $c_5$-type terms have been calculated from the results for 
baryon masses in Ref.~\cite{Frink:2004ic},  while the ones for $c_6$- and $c_7$-type
constants use the analysis of hyperon form factors in Ref.~\cite{Kubis:BaryonFF}. 

We finally turn to the $S=2$ sector.
As the $\Xi^0$ and the $\Xi^-$ are an isospin doublet, the Lagrangian for the 
$\pi\Xi$ system can be copied immediately from $\Lagr_{\pi N}^{(2)}$,
\begin{align}
\Lagr_{\pi \Xi}^{(1+2)}&= \bar \Xi \biggl\{ i \slashed{D} - \krig{m}_\Xi +\frac{g_\Xi}{2}\slashed{u}\gamma_5
+c_1^\Xi \langle \chi_+ \rangle
-\frac{c_2^\Xi}{8\krig{m}_\Xi^2} \bigl( \langle u_\mu u_\nu \rangle \{D^\mu,D^\nu\} + {\rm h.c.} \bigr)
+\frac{c_3^\Xi}{2} \langle u_\mu u^\mu\rangle\nn
&\quad+ \frac{\,c_4^\Xi}{4}[u_\mu,u_\nu] i\sigma^{\mu\nu}
+ c_5^\Xi \tilde \chi_-
+ \frac{1}{8\krig{m}_\Xi} \bigl( c_6^\Xi \F_{\mu\nu}^+ 
+ c_7^\Xi \langle \F_{\mu\nu}^+ \rangle \bigr) \sigma^{\mu\nu} \biggr\} \Xi ~,
\end{align}
where we have used $\F_{\mu\nu} = (\partial_\mu A_\nu-\partial_\nu A_\mu) 
\mathcal{Q}_\Xi$, $\mathcal{Q}_\Xi=e\,\mathrm{diag}(0,-1)$. 
The matching relations read
\begin{align}
\krig{m}_\Xi &= m_0 -4M_K^2 \bigl(b_0+b_D+b_F\bigr) 
- \frac{M_K^3}{48\pi \Fpi^2}\biggl[5D^2+6DF+9F^2+\frac{4}{3\sqrt{3}}(D+3F)^2\biggr]
+ \Order(M_K^4) ~, \nn
g_\Xi &= D-F + \Order(M_K^2) ~, \quad
c_1^\Xi =b_0+\frac{b_D}{2}-\frac{b_F}{2}  
+ \frac{M_K}{256\pi \Fpi^2}\bigg[5D^2+6DF+9F^2+\frac{2}{3\sqrt{3}}(D+3F)^2\bigg]
+ \Order(M_K^2) ~,\nn
c_2^\Xi&= b_8-b_9+b_{10}+ 2b_{11} - \frac{M_K}{128 \pi F_\pi^2}\bigg[
6 + \frac{19}{3} D^4 + \frac{58}{3} D^2 F^2  + 25 F^4 - 4 DF(D^2-3F^2) 
+\frac{8(D-F)^2 (D+3 F)^2}{3\sqrt{3}} \bigg]\nn
& \qquad +\Order(M_K^2) ~,\nn
c_3^\Xi&=b_1-b_2+b_3+2b_4+\frac{M_K}{128 \pi F_\pi^2 }\bigg[
5D^2 + 6DF + 9F^2
+ \frac{19}{3} D^4 + \frac{58}{3} D^2 F^2  +  25 F^4 - 4DF(D^2-3F^2)  \nn
&\qquad +\frac{8 (D-F)^2 (D+3 F)^2}{3\sqrt{3}}
\bigg]+\Order(M_K^2) ~,\nn
c_4^\Xi&=4(b_{6}-b_{5}) - \frac{M_K}{96 \pi F_\pi^2}\bigg[
D^2 +6 DF - 3F^2
- \frac{9}{2} D^4 + 10 D^3 F +  D^2 F^2 + 18 D F^3 - \frac{33}{2} F^4 
- \frac{2 (D-F)^2 (D+3 F)^2}{\sqrt{3}}\bigg]\nn
&\qquad +\Order(M_K^2) ~,\nn
c_5^\Xi &= -b_D+b_F + \frac{M_K}{128\pi \Fpi^2}\bigg[D^2+6DF-3F^2+\frac{8}{3\sqrt{3}}(D+3F)(D-F)\bigg] +\Order(M_K^2) ~, \nn
\frac{c_6^\Xi}{\krig{m}_\Xi} &= 8\bigl( b_{12}- b_{13}\bigr) + \frac{M_K}{24\pi \Fpi^2}(D^2+6DF-3F^2)   +\Order(M_K^2)  ~, \quad
\frac{c_7^\Xi}{\krig{m}_\Xi} = \frac{16}{3}b_{13} -\frac{M_K}{8\pi \Fpi^2}(D+F)^2   +\Order(M_K^2) ~. \label{eq:matchX}
\end{align}
\end{widetext}
If one evaluates the $\Order(M_K)$ shifts in the matching relations~\eqref{eq:matchL}, \eqref{eq:matchS}, \eqref{eq:matchX} 
one in general finds sizeable shifts in particular in the $c_2$- and $c_3$-type couplings, 
which tend to cancel to some extent in the combinations featuring in 
the pion--hyperon scattering lengths.

\subsection{Threshold amplitudes in SU(2) and  low-energy theorems}

An interesting application of the above results is the representation of 
the pion--hyperon threshold amplitudes in chiral SU(2).
As it will turn out below, these can be used to prove low-energy theorems
for certain of these amplitudes, comparable to the well-known one 
for the isovector pion--nucleon scattering length, 
or the isovector pion--kaon scattering length proven similarly in chiral SU(2) for kaons~\cite{Roessl}.
Although experimental verifications of these pion--hyperon low-energy theorems may be very difficult,
they may become testable on the lattice soon.

In pion--nucleon scattering,
isospin-even and -odd amplitudes are given in terms of those of definite isospin as
\begin{align}
T^+(\nu,t) &= \frac{1}{3}\bigl(T^{(1/2)}+2T^{(3/2)}\bigr) ~,
\nn T^-(\nu,t) &= \frac{1}{3}\bigl(T^{(1/2)}-T^{(3/2)}\bigr) ~,
\end{align}
where $\nu=s-u$. $T^\pm(\nu,t)$ are even/odd under crossing $\nu \leftrightarrow -\nu$, leading to polynomial parts of the form
\begin{align}
T^+_{\rm pol}(\nu,t) &= \alpha_0^+ + \alpha_1^+ t + \alpha_2^+ \nu^2 + \alpha_3^+ t^2 + \ldots ~,\nn
T^-_{\rm pol}(\nu,t) &= \nu\bigl(\alpha_0^- + \alpha_1^- t + \alpha_2^- \nu^2 + \ldots \bigr) ~.
\end{align}
At threshold, $\nu_{\rm thr} = 4m_N M_\pi$, $t_{\rm thr} = 0$, therefore
\begin{align}
T^+_{\rm pol}(\nu_{\rm thr},0) = \sum_{i=1}^\infty \gamma_i^+ M_\pi^{2i} ~, \quad
\nn T^-_{\rm pol}(\nu_{\rm thr},0) = \sum_{i=1}^\infty \gamma_i^- M_\pi^{2i-1} ~.\label{eqn:TpiN}
\end{align}
In particular, $\gamma_1^- = 1/(2\Fpi^2)$~\cite{Weinberg}. 
Deviations from this scheme, e.g.\ $\Order(M_\pi^3)$ contributions in $T_{\pi N}^+$, are nonanalytic and due to pion loops. 
The important point about the low-energy theorem for $\gamma_1^-$ is that it
does not contain any unknown constants, hence in an SU(3) calculation of this
quantity, it does not get renormalized by large kaon-mass effects. 
Corrections to $a_{\pi N}^-$ are of $\Order(M_\pi^3)$ even then and should remain moderate in size.

The well-known pion--nucleon threshold amplitudes up to $\Order(M_\pi^3)$ read
\begin{align}
T_{\pi N}^+&= \frac{M_\pi^2}{\Fpi^2}\biggl\{-\frac{g^2}{4 m_N}
+2(c_2+c_3-2c_1) +\frac{3 g^2 M_\pi}{64\pi \Fpi^2} 
\nn&\qquad
+\Order\big(M_\pi^2\big) \biggr\} ~,\nn
T_{\pi N}^-&=\frac{M_\pi}{2 \Fpi^2} \biggl\{1+\frac{g^2 M_\pi^2}{4m_N^2}+\frac{M_\pi^2}{8\pi^2\Fpi^2}
\Big(1-2\log\frac{M_\pi}{\mu} \Big)
\nn&\qquad
 + M_\pi^2 d_{\pi N}^r(\mu) +\Order\big(M_\pi^4\big) \biggr\} ~.\label{eq:piNthresh}
\end{align}
The vanishing of the term of $\Order(M_\pi^4)$ in $T_{\pi N}^-$ was found in Ref.~\cite{BKM:Mpi4}.
We have not included  $\Order(q^3)$ counterterms in our SU(3) calculation, and therefore refrain
from giving matching relations at that order.
We only add a generic $\Order(q^3)$ counterterm $\propto d_{\pi N}(\mu)$ to the isovector scattering length
that cancels divergence and scale dependence of the loop contributions 
(in Ref.~\cite{BKM:Mpi4}, this generic LEC was called $B^r (\mu)$).
Of course in the case of $\pi N$ scattering, this combination is known~\cite{BKM93,FMS98},
\beq
d_{\pi N}^r = 8\big(d^{\rm r}_1+d^{\rm r}_2+d^{\rm r}_3+2d^{\rm r}_5\big) 
~\bigg[\!+\frac{2}{\Fpi^2}l_4^r\bigg] ~
\eeq
(where the inclusion of the term proportional to $l_4^r$ depends on the convention of the Lagrangians
used~\cite{Ecker:1995rk,GR02}).  We only remark that kaon loops contribute a
term to the matching of $d_{\pi N}^r$ according to
\beq
d_{\pi N}^r = - \frac{1}{16\pi^2\Fpi^2} \Big(1+2\log\frac{M_K}{\mu} \Big) + \ldots ~. 
\label{eq:dpiN_Kloop}
\eeq

Next we turn to the SU(2) representations of the pion--hyperon scattering
amplitudes at threshold. Isospin and crossing symmetry dictate the
following structure of the $\pi\Sigma$ scattering amplitude:
\begin{align}
T\bigl(\pi^i\Sigma^a\to\pi^j\Sigma^b\bigr) &=
\bar T_{\pi\Sigma}( \nu,t) \delta^{ai}\delta^{bj}
+ \bar T_{\pi\Sigma}(-\nu,t) \delta^{aj}\delta^{bi} \nn
&\quad+ \tilde T_{\pi\Sigma}(\nu,t) \delta^{ab}\delta^{ij} ~. 
\end{align}
These are related to amplitudes of definite isospin by 
\begin{align}
T_{\pi\Sigma}^{(0)} &= 3\bar T_{\pi\Sigma}(\nu,t) + \bar T_{\pi\Sigma}(-\nu,t) + \tilde T_{\pi\Sigma}(\nu,t) ~, \nn
T_{\pi\Sigma}^{(1)} &= \tilde T_{\pi\Sigma}(\nu,t) - \bar T_{\pi\Sigma}(-\nu,t)  ~, \nn
T_{\pi\Sigma}^{(2)} &= \tilde T_{\pi\Sigma}(\nu,t) + \bar T_{\pi\Sigma}(-\nu,t)  ~.
\end{align}
The combinations of amplitudes \emph{odd} and \emph{even} in $\nu$ (analogous to $T^\pm_{\pi N}$) are therefore
\begin{align}
\bar T_{\pi\Sigma}^-(\nu,t):\!&=\bar T_{\pi\Sigma}(\nu,t) - \bar T_{\pi\Sigma}(-\nu,t)  
\nn&= \frac{1}{3}T_{\pi\Sigma}^{(0)} + \frac{1}{2}T_{\pi\Sigma}^{(1)} - \frac{5}{6}T_{\pi\Sigma}^{(2)} ~,\nn
\bar T_{\pi\Sigma}^+(\nu,t):\!&=\bar T_{\pi\Sigma}(\nu,t) + \bar T_{\pi\Sigma}(-\nu,t)  \nn
&= \frac{1}{3}T_{\pi\Sigma}^{(0)} - \frac{1}{2}T_{\pi\Sigma}^{(1)} + \frac{1}{6}T_{\pi\Sigma}^{(2)} ~,
\end{align}
and $\tilde T_{\pi\Sigma}(\nu,t)$ is also even. At threshold, where $\nu_{\rm thr}=4m_\Sigma M_\pi$, our results read
\begin{align}
\tilde T_{\pi\Sigma}&=\frac{M_\pi^2}{\Fpi^2}\biggl\{ -\frac{g_\Sigma^2}{4 m_\Sigma} + 4\big(2 c_{2a}^{\Sigma}+c_{3a}^{\Sigma}-c_1^{\Sigma}\big)(\mu)
\nn&
+\frac{3g_{\Sigma\Lambda}^2 M_\pi}{64\pi \Fpi^2}f\left(m_\Sigma\!-\!m_\Lambda,M_\pi,\mu\right) 
+\Order\big(M_\pi^2\big) \biggr\} ~, \nn
\bar T_{\pi\Sigma}^+&=\frac{M_\pi^2}{\Fpi^2}\biggl\{ \frac{g_\Sigma^2}{4 m_\Sigma} - \frac{g_{\Sigma\Lambda}^2}{2(m_\Lambda+ m_\Sigma)}
+ 4\big(4 c_{2b}^{\Sigma}+c_{3b}^{\Sigma}\big)(\mu)
\nn&
+\frac{3[g_\Sigma^2-g_{\Sigma\Lambda}^2f\left(m_\Sigma\!-\!m_\Lambda,M_\pi,\mu\right)] M_\pi}{32\pi\Fpi^2} +\Order\big(M_\pi^2\big) \biggr\} ~,\nn
\bar T_{\pi\Sigma}^-&=\frac{2 M_\pi}{\Fpi^2}\biggl\{ 1
+\frac{g_\Sigma^2M_\pi^2}{16 m_\Sigma^2}
+\frac{g_{\Sigma\Lambda}^2 M_\pi^2}{4(m_\Lambda+m_\Sigma)^2}+ M_\pi^2 d_{\pi\Sigma}^r(\mu)
\nn&
+\frac{M_\pi^2}{8\pi^2\Fpi^2}\Big(1-2\log\frac{M_\pi}{\mu}\Big) 
 +\Order\big(M_\pi^3\big)\biggr\} ~, \label{eq:piSigma}
\end{align}
where
\begin{align}
f\left(\delta,M_\pi,\mu\right) &= 
\frac{2}{\pi}\sqrt{1-\frac{\delta^2}{M_\pi^2}} \arccos\Big(-\frac{\delta}{M_\pi}\Big) \nonumber \\
& \quad + 
\frac{\delta}{\pi M_\pi} \bigg(2\log\frac{M_\pi}{\mu} - \frac{1}{3}\bigg) ~.\label{eq:f}
\end{align}
The indicated dependence of the constants $c_i^\Sigma$ on the renormalization scale $\mu$ compensates the one
from $f\left(m_\Sigma-m_\Lambda,M_\pi,\mu\right)$, where the corresponding $\beta$-functions
are suppressed by the small parameter $m_\Sigma-m_\Lambda$.
We find that there exists a similar low-energy theorem for $\bar T_{\pi\Sigma}^-$ at threshold
as for $T_{\pi N}^-$: the leading term in the pion-mass expansion is fixed simply by $F_\pi$, 
and corrections are suppressed by a relative factor of $M_\pi^2$.
Note that the $\Lambda-\Sigma$ transition generates terms of order $M_\pi^4$
in $\bar T_{\pi\Sigma}^-$ in contrast to the pion--nucleon case. 
Up to order $M_\pi^3$, however, there is no nonanalytic dependence of $T_{\pi\Sigma}^-$
on $m_\Sigma-m_\Lambda$: such terms, as encoded in the function $f(m_\Sigma-m_\Lambda, M_\pi,\mu)$ 
only feature in the iso\emph{scalar} scattering lengths in \eqref{eq:piSigma}.

The $\pi\Lambda$ scattering amplitude
\begin{equation}
T\bigl(\pi^i\Lambda\to\pi^j\Lambda) = T_{\pi\Lambda}(\nu,t) \delta^{ij}
\end{equation}
is even under crossing $\nu \leftrightarrow -\nu$ and therefore, according to
the arguments in the preceding section, starts at $\Order(M_\pi^2)$ at threshold. 
This is obvious also from a different point of view: no Weinberg--Tomozawa
term exists. Thus the result reads
\begin{align}
T_{\pi \Lambda}&=\frac{M_\pi^2}{\Fpi^2}\biggl\{-\frac{g_{\Sigma\Lambda}^2}{2(m_\Lambda+m_\Sigma)} 
+ 2\big(c^\Lambda_2+c^\Lambda_3-2c^\Lambda_1\big)(\mu)
\nn&
+\frac{3g_{\Sigma\Lambda}^2M_\pi}{64\pi\Fpi^2}f\left(m_\Lambda\!-\!m_\Sigma,M_\pi,\mu\right) +\Order\big(M_\pi^2\big) \biggr\} ~,
\end{align}
with $f\left(m_\Lambda-m_\Sigma,M_\pi,\mu\right)$ as defined in \eqref{eq:f}.

Finally, from the point of view of isospin, crossing etc., $\pi\Xi$ scattering is identical to 
the $\pi N$ case, hence the \emph{odd} and \emph{even} combinations of amplitudes are 
identical to those in~\eqref{eqn:TpiN}. 
The result of the calculation is also formally equal to the one from the
pion--nucleon case and reads
\begin{align}
T_{\pi\Xi}^+&= \frac{M_\pi^2}{\Fpi^2}\biggl\{-\frac{g_\Xi^2}{4 m_\Xi}
+2(c^\Xi_2+c^\Xi_3-2c^\Xi_1) +\frac{3 g_\Xi^2 M_\pi}{64\pi \Fpi^2}  \nn&\qquad
+\Order\big(M_\pi^2\big) \biggr\} ~,\nn
T_{\pi\Xi}^-&=\frac{M_\pi}{2 \Fpi^2} \biggl\{1+\frac{g_\Xi^2 M_\pi^2}{4m_\Xi^2}+\frac{M_\pi^2}{8\pi^2\Fpi^2}
\Big(1-2 \log \frac{M_\pi}{\mu}\Big) \nn&\qquad
+ M_\pi^2 d_{\pi\Xi}^r(\mu) +\Order\big(M_\pi^4\big)\biggr\} ~.
\end{align}
Again, $T_{\pi\Xi}^-$ obeys the same type of low-energy theorem as $T_{\pi N}^-$.
We remark that the $\Order(q^3)$ couplings $d_{\pi\Sigma}^r$, $d_{\pi\Xi}^r$ receive
the same kaon-loop contributions when matching to SU(3) as $d_{\pi N}^r$ in~\eqref{eq:dpiN_Kloop}.

\subsection{Chiral extrapolations}

We wish to comment on the consequences of the low-energy theorems for the isospin-odd
threshold amplitudes $T_{\pi N}^-$, $\bar T_{\pi \Sigma}^-$, and $T_{\pi\Xi}^-$, for 
studies of these quantities on the lattice, where typically pion masses larger than
the physical ones are employed.  

If we take into account the pion-mass dependence of $F_\pi$,
\beq
\Fpi=F\left\{1+\frac{M_\pi^2}{F^2}\left(l_4^r-\frac{1}{8\pi^2}\log \frac{M_\pi}{\mu}\right)\right\}+\Order\big(M_\pi^4\big)
~,
\eeq
where $F$ is the pion decay constant in the SU(2) chiral limit, $T_{\pi N}^-$ can be rewritten from
\eqref{eq:piNthresh} in the form
\beq
T_{\pi N}^- = \frac{M_\pi}{2 F^2} \Big\{ 1+ M_\pi^2 d_{\pi N}^{\rm eff} + \Order(M_\pi^4) \Big\} ~,\label{eq:aMinus}
\eeq
hence there is no $M_\pi^3 \log M_\pi$ term.
The low-energy constant $l_4^r$ can be determined from a dispersive analysis of the scalar form 
factor of the pion~\cite{CGL}, leading to $F=(86.9\pm0.2)$\,MeV.  
With this information, we can fix the effective coupling $d_{\pi N}^{\rm eff}$ from 
the experimental information on the isovector $\pi N$ scattering length~\cite{MRR},
$a_{\pi N}^-=(85.2\pm 1.8)\times10^{-3}M_\pi^{-1}$, leading to $d_{\pi N}^{\rm eff} = (-2.4\pm 1.0)\,{\rm GeV}^{-2}$,
which is perfectly of natural order.

\begin{figure}
\begin{center}\includegraphics[width=\linewidth]{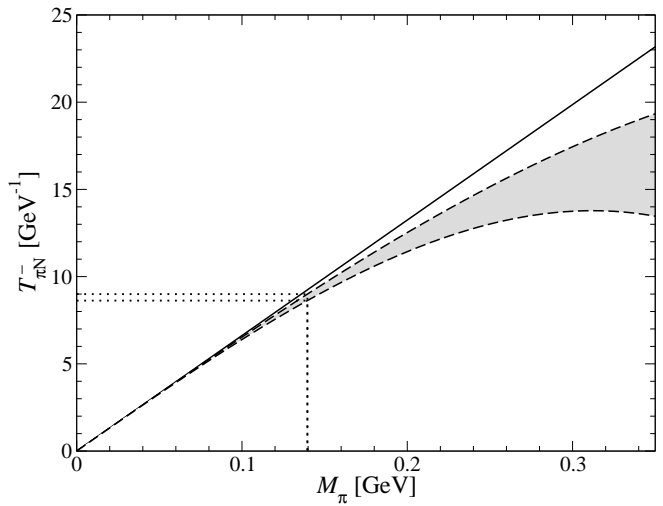}\end{center}
\caption{The threshold amplitude $T_{\pi N}^-$ for varying pion mass $0\leq M_\pi\leq 350$~MeV.
The full line denotes the parameter-free leading-order result, 
the dashed band is the expression valid up-to-and-including $\Order(M_\pi^4)$, fitted
to the experimental value (marked by dotted lines) extracted in Ref.~\cite{MRR} 
at the physical pion mass.\label{fig:aMinus}}
\end{figure}
In Fig.~\ref{fig:aMinus}, we compare the pion-mass dependence of $T_{\pi N}^-$, with $d_{\pi N}^{\rm eff}$
varying within its error range, to the parameter-free current algebra prediction.
At $M_\pi=300\,$MeV, the corrections are still moderate, about $(-20\pm10)\%$.  
From dimensional arguments, one would expect the omitted $\Order(M_\pi^5)$ terms to contribute less than 
5\% at such masses.  
$T_{\pi N}^-$ is therefore a very stable quantity under chiral corrections.  This is in contrast to
$T_{\pi N}^+$, which is afflicted by not very precisely known counterterms already at leading order.
We wish to remark again that the omission of explicit dynamical $\Delta(1232)$ contributions
does not necessarily invalidate the chiral extrapolation for S-wave scattering lengths: 
as can be checked from the explicit expressions in Ref.~\cite{Fettes:2000bb}, 
the effects of $\pi\Delta$ intermediate states are entirely smooth and can be well approximated 
by tadpole-type loops, with the $\Delta$ propagator shrunk to a point.
In fact, there are no $\pi\Delta$ loop contributions to $T_{\pi N}^-$ up to $\Order(\epsilon^3)$
in the small-scale expansion.
Compare also the smooth pion-mass dependence of the nucleon and $\Delta$ masses in Ref.~\cite{Hoja}.

For lack of data to fix the respective $\Order(M_\pi^3)$ constants $d_{\pi\Sigma}^{\rm eff}$,  $d_{\pi\Xi}^{\rm eff}$, 
we refrain from showing chiral extrapolations for $\bar T_{\pi \Sigma}^-$ and $T_{\pi\Xi}^-$ graphically.
Their pion-mass dependence is completely analogous to~\eqref{eq:aMinus},
\begin{align}
\bar T_{\pi \Sigma}^- &= \frac{2M_\pi}{F^2} \Big\{ 1+ M_\pi^2 d_{\pi\Sigma}^{\rm eff} + \Order(M_\pi^3) \Big\} ~,\nn
T_{\pi \Xi}^- &= \frac{M_\pi}{2 F^2} \Big\{ 1+ M_\pi^2 d_{\pi\Xi}^{\rm eff} + \Order(M_\pi^4) \Big\} ~,\label{eq:apiSigmaXi}
\end{align}
and can be expected to be similarly stable. [One caveat in the case of
the $\pi\Sigma$ scattering length is that it becomes complex for very small
pion masses, when the $\Sigma$ becomes unstable against strong decays
into $\pi\Lambda$.  However, for practical purposes (i.e., chiral
extrapolations in the context of lattice simulations), problems with pion masses
less than half the physical ones seem somewhat academic at present.]  
We wish to point out once more that the relations~\eqref{eq:apiSigmaXi} are a consequence of chiral SU(2)
symmetry: an exploitation of three-flavor ChPT does not guarantee that
there are no corrections of $\Order(M_K^{2n})$ to the leading term.

\begin{acknowledgments}
We would like to thank Matthias Frink for useful discussions, in particular on issues
of Lagrangian construction. We are grateful to Silas Beane and Martin Savage
for drawing our attention to Ref.~\cite{Torok}.
This work is supported in part by the European Community-Research Infrastructure
Integrating Activity ``Study of Strongly Interacting Matter''
(acronym HadronPhysics2, Grant Agreement No.~227431), by Deutsche Forschungsgemeinschaft 
(SFB/TR 16, ``Subnuclear Structure of Matter'') and  by the Helmholtz Association through
funds provided to the virtual institute ``Spin and strong QCD'' (VH-VI-231).
\end{acknowledgments}

\begin{widetext}

\appendix

\section{One-loop integrals}
\subsection{Threshold expansion}\label{App:thr-exp}
\subsubsection{One-point functions}

In the purely mesonic sector one finds, using dimensional regularization
\begin{align}
I_M(M^2_X) :\!&=\frac{1}{i}\int_{\IR} \frac{d^dk}{(2\pi)^d}\frac{1}{M_X^2-k^2}=2M_X^2
\bigg(\lambda+\frac{1}{16\pi^2}\log{\frac{M_X}{\mu}} \bigg),\\
\lambda &=\frac{\mu^{d-4}}{(4\pi)^{d/2}}\bigg(\frac{1}{d-4}-\frac{1}{2}\big(\log{4\pi}-\gamma+1\big)\bigg).\nonumber
\end{align}
The subscript $\IR$ means that the infrared-regular part is neglected. 
$\mu$ is the dimensional regularization scale and $M_X$ the mass of the meson involved. 
Here and in the following, we suppress the infinitesimal imaginary parts $-i\epsilon$ in all propagators for the sake of brevity. 

\subsubsection{Two-point functions}

The two-point function with two meson propagators can be calculated in closed form.
We use the following combinations of momenta relevant for the calculation at hand:
\beq
\Sigma_\mu=(p_a+q_i)_\mu=(p_b+q_j)_\mu ~,\quad
\Delta_\mu=(q_j-q_i)_\mu=(p_a-p_b)_\mu ~,\quad
Q_\mu=(p_a+p_b)_\mu ~. 
\eeq
For the two-meson scalar loop integral one obtains 
\begin{align}
I_{MM}(t,M^2_X,M^2_Y):\!&=\frac{1}{i}\int_{\IR}\frac{d^dk}{(2\pi)^d}\frac{1}{(M_X^2-k^2)(M_Y^2-(k-\Delta)^2)}\nn
&= -2\lambda-\frac{1}{32\pi^2 t}\Biggl\{
t\biggl(-2+\log{\frac{M_X^2}{\mu^2}}+\log{\frac{M_Y^2}{\mu^2}}\biggr)+(M_X^2-M_Y^2)\log{\frac{M_X^2}{M_Y^2}}\nn
&\qquad\qquad\qquad\qquad-\sqrt{\lambda(M_Y^2,M_X^2,t)}\artanh\Biggl(\frac{\sqrt{\lambda(M_Y^2,M_X^2,t)}}{-t+(M_X+M_Y)^2}\Biggr)
\Biggr\}~,
\end{align}
where we use K\"all\'en triangle function
$\lambda(a,b,c):=a^2+b^2+c^2-2(ab+bc+ac)$ and $t:=\Delta^2$.
The representation above is useful and unambiguous for $t\leq 0$.

In the one-meson-one-baryon function both the infrared-regular and
-singular part are nonvanishing. 
The infrared-singular part can be expressed in terms of
elementary functions. 
For arbitrary $p^2$ the result reads
\begin{align}
I_{MB}(p^2,M^2_X):\!&=\frac{1}{i}\int_{\IR}\frac{d^dk}{(2\pi)^d}\frac{1}{(k^2-M_X^2)((k-p)^2-m_0^2)}\nn
&=\frac{m_0^2-M_X^2-p^2}{p^2}\bigg(\lambda_X-\frac{1}{32 \pi ^2}\bigg) 
-\frac{\sqrt{-\lambda(p^2,m_0^2,M_X^2)}}{16 \pi ^2 p^2} 
\arccos\bigg(\frac{m_0^2-M_X^2-p^2}{2 M_X \sqrt{p^2}}\bigg) ~,\label{eqn:IMBgen}
\end{align}
where in our calculation $I_{MB}$ features with the arguments $p^2=\{s,u,m_0^2\}$.
$\lambda_Y$ is defined as
\begin{equation}
\lambda_Y:=\lambda+\frac{1}{16\pi^2}\log\Big({\frac{M_Y}{\mu}}\Big).
\end{equation}

\subsubsection{Three-point functions}

The three-point function with two mesons involved can be traced back to the
one-meson-one-baryon function, since the combination of two mesons due
to Feynman parameter integration corresponds to a new structure similar
to a meson propagator,
\begin{align}
I_{MMB}(t,M^2_X,M^2_Y):\!&=\frac{1}{i}\int_{\IR}\frac{d^dk}{(2\pi)^d}\frac{1}{(M_X^2-k^2)(M_Y^2-(k-\Delta)^2)(m_0^2-(p_i-k)^2)}
\nn
&= \frac{1}{i}\int_0^1dz\int_{\IR}
\frac{d^dk}{(2\pi)^d}\frac{1}{[k^2-\mathcal{M}(z)^2]^2(m_0^2-(p_i-k-z\Delta)^2)}
\nn
&= \int_0^1dz\Bigl( -\frac{\partial}
{\partial\mathcal{M}(z)^2}\Bigr)I_{MB}\big((p_i-z\Delta)^2,\mathcal{M}(z)^2\big)~,
\label{eqn:Immb}
\end{align}
with $\mathcal{M}^2(z)=zM_Y^2+(1-z)M_X^2 - z(1-z)\Delta^2$.
A closer look at the Passarino--Veltman (PaVe) reduction of the only topology to which such
integrals contribute, diagram (l) in Fig.~\ref{pic:graphs}, shows that the
corresponding denominators are nonvanishing at threshold, thus we
require only the zeroth order of the integral expansion around the threshold. 
For equal meson masses and $t=0$, $\mathcal{M} = M_X$ and we can easily evaluate~\eqref{eqn:Immb}. 
The  result reads
\begin{align}
I_{MMB}(0,M^2_X,M^2_X):\!&=\frac{1}{m_0^2}\Big(\lambda_X+\frac{1}{32\pi^2}\Big)+\frac{2m_0^2-M_X^2}{16\pi^2m_0^2M_X\sqrt{4m_0^2-M_X^2}}\arccos\bigg(-\frac{M_X}{2m_0}\bigg)~.
\end{align}
In the case of different meson masses we can rewrite the derivative according to the chain rule
\beq
I_{MMB}(0,M^2_X,M^2_Y) =\frac{1} {M_X^2-M_Y^2}\int_0^1dz\Bigl(\frac{\partial}{\partial z}\Bigr)
I_{MB}\big(p_i^2,\mathcal{M}(z)^2\big)
=\frac{I_{MB}(p_i^2,M_Y^2)-I_{MB}(p_i^2,M_X^2)}{M_X^2-M_Y^2}~,
\eeq
which is of course the representation of the derivative in~\eqref{eqn:Immb} in the limit $M_X \rightarrow M_Y$.

The three-point function with two baryon propagators reads
\beq
I_{MBB}(P_1^2,P_2^2,M^2_X):=
\frac{1}{i}\int_{\IR}\frac{d^dk}{(2\pi)^d}\frac{1}{(M_X^2-k^2)(m_0^2-(k-P_1)^2)(m_0^2-(k-P_2)^2)}~.
\eeq
It is easy to see from the graphs in Fig.~\ref{pic:graphs} that at least one of the momenta $P_{1/2}$
is always on its mass shell, so without loss of generality we choose $P_1^2=m_0^2$.
Furthermore, there are two possible momentum configurations to consider:
$P_2^2 = \{m_0^2,s\}$ (the crossed channel $P_2^2=u$ will be discussed below), 
together with $q^2:=(P_2-P_1)^2 = \{t,M^2\}$, where $M$ is the mass of one of the external mesons.
$q^2$ is a quantity of second chiral order in both cases.

Applying the method of infrared regularization, the above integral reads 
\begin{align}
I_{MBB}&=\frac{\Gamma(3-d/2)}{(4\pi)^{d/2}m^{d-4}}\int^{\infty}_0dx\int^{\infty}_0dy\Big[x^2m_0^2+y^2P_2^2
+xy(P_2^2+m_0^2-q^2)+y(m_0^2-M_X^2-P_2^2)+M_X^2-xM_X^2\Bigr]^{d/2-3}\nn
&=\frac{\Gamma(3-d/2)}{(4\pi)^{d/2}}\frac{M_X^{d-4}}{m_0^{d-5}\sqrt{P_2^2}}\int^{\infty}_0d\tilde
x\int^{\infty}_0d\tilde y\Big[(\tilde x+\tilde y)^2
-\tilde x\tilde y\frac{\epsilon}{m_0\sqrt{P_2^2}}+\tilde
y\frac{m_0^2-M_X^2-P_2^2}{\sqrt{P_2^2}M_X}+1-\tilde
x\frac{M_X}{m_0}\Bigr]^{d/2-3},
\end{align}
where we have substituted
\beq
y=\tilde y \frac{M_X}{\sqrt{P_2^2}} ~, \quad x=\tilde x \frac{M_X}{m_0} ~, \quad
\epsilon := q^2-\Big(\sqrt{P_2^2}-m_0\Big)^2 ~.
\eeq
The variable $\epsilon$ is chosen in such a way that it is small around
and vanishing at threshold, becoming our small expansion parameter.
Because the PaVe reduction procedure yields denominators linear in
$\epsilon$, we evaluate the integrals  up to and including first
order in $\epsilon$. After a transformation of the variables as $\tilde
x:=u/2$ and $\tilde y := (2u-v)/2$ we obtain
\begin{align}
I_{MBB}&=\frac{\Gamma(3-d/2)}{2(4\pi)^{d/2}}\frac{M_X^{d-4}}{m_0^{d-5}\sqrt{P_2^2}}\int^{\infty}_0du\int^{2u}_0dv\Big[
u^2+1-u\frac{P_2^2+M_X^2-m_0^2}{\sqrt{P_2^2}M_X}\nn
& \qquad\qquad\qquad\qquad\qquad\qquad\qquad\qquad\quad
+v\frac{P_2^2-m_0^2+M_X^2(1-\sqrt{P_2^2}/m_0)}{2\sqrt{P_2^2}M_X^2}-\epsilon
\frac{2uv-v^2}{4m_0\sqrt{P_2^2}}
\Bigr]^{d/2-3} \nn
&=\frac{\Gamma(3-d/2)}{2(4\pi)^{d/2}}\frac{M_X^{d-4}}{m_0^{d-5}\sqrt{P_2^2}}\int^{\infty}_0du\int^{2u}_0dv\bigg\{\Big[u^2+1-u\frac{P_2^2+M_X^2-m_0^2}{\sqrt{P_2^2}M_X}+Kv\Bigr]^{d/2-3}
\\
&\qquad\qquad\qquad\qquad\qquad\qquad\qquad\qquad\quad
+\epsilon\frac{d-6}{2}\frac{v^2-2uv}{4m_0\sqrt{P_2^2}}\Big[u^2+1-u\frac{P_2^2+M_X^2-m_0^2}{\sqrt{P_2^2}M_X}+Kv\Bigr]^{d/2-4}\bigg\}
+\Order(\epsilon^2) ~, \nonumber
\end{align}
where in the last step we have expanded the integrand to first order in $\epsilon$ and defined
\beq
K:=\frac{P_2^2-m_0^2+M_X^2(1-\sqrt{P_2^2}/m_0)}{2\sqrt{P_2^2}M_X}~. 
\eeq
Integrating over $v$ one obtains the following combination of one-parameter integrals
\begin{align}
I_{MBB}=\frac{\Gamma(3-d/2)}{2(4\pi)^{d/2}}&\frac{M_X^{d-4}}{m_0^{d-5}\sqrt{P_2^2}}\bigg\{
K^{-1}\frac{2}{d-4}(\mathcal{D}^{2,0}-\mathcal{A}^{2,0}) 
\nn&\qquad\qquad\quad
+\epsilon\frac{d-6}{8m_0\sqrt{P_2^2}}\frac{4(\mathcal{D}^{1,0}-\mathcal{A}^{1,0})-2(d-2)K
(\mathcal{A}^{2,1}+\mathcal{D}^{2,1})}{(d^3-12 d^2+44 d-48)K^3}\bigg\}
+\Order(\epsilon^2) ~,
\end{align}
where we denote the one-parameter integrals as follows
\begin{align}
\mathcal{A}^{Z,y} &:=\int^\infty_0 u^y[(u-c_0)^2+C_0]^{d/2-Z}du  &\text{with}&
&C_0&=1-c_0^2 ~, & c_0 &=\frac{P_2^2+M_X^2-m_0^2}{2M_X\sqrt{P_2^2}}~, \nn
\mathcal{D}^{Z,y} &:=\int^\infty_0 u^y[(u-d_0)^2+D_0]^{d/2-Z}du  &\text{with}&
&D_0&=1-d_0^2 ~, & d_0 &=\frac{M_X}{m_0}~.  \label{eqn:AF}
\end{align}
The parameter integrals of the type $\mathcal{D}^{Z,y}$ do not cause problems
for any of the combinations $\{ Z,y \}$. The $\mathcal{A}$-type
integrands on the other hand may have poles within the integration
range, where they only converge for $d<4$ and thus analytic continuation
must be performed.  The results of this are given in Appendix~\ref{App:analytic-continuation}. 

For the crossed channel the
steps above remain valid when substituting $s\rightarrow u$, and 
for $P_2^2 \in \{m_0^2,u\}$ the expansion parameter yields
$\epsilon:=q^2-(\sqrt{P_2^2}+m_0)^2$. The final results are presented in
Appendix~\ref{App:Integrals-results}.

\subsubsection{Four-point function}

This integral only appears in the box graph (i) of Fig.~\ref{pic:graphs} and has only one 
momentum structure each in $s$- and $u$-channel,
$\{P_1^2=P_3^2=m_0^2,\,P_2^2=s\}$ and $\{P_1^2=P_3^2=m_0^2,\,P_2^2=u\}$ respectively,
\beq
I_{MBBB}(P_2^2,M_X^2):= \frac{1}{i}\int_{\IR}\frac{d^dk}{(2\pi)^d}\frac{1}{(M_X^2-k^2)(m_0^2-(k-P_1)^2)(m_0^2-(k-P_2)^2)(m_0^2-(k-P_3)^2)}~.
\eeq
The threshold expansion parameter can here be chosen to be the
Mandelstam variable $t$, which is also a quantity of  second chiral
order. When we fix the remaining Mandelstam variables to their values at the
threshold, the PaVe denominators become linear functions in $t$, which demands
an expansion of the scalar loop integrals  up to $\Order(t)$.

Combining the baryon propagators with the meson propagator in the
spirit of infrared regularization and performing the loop
integration we find
\begin{align}
I_{MBBB}&:=\frac{\Gamma(4-d/2)}{(4\pi)^{d/2}}\int^\infty_0 dx\int^\infty_0 dy\int^\infty_0 dz \nn
\times \bigg\{&\Bigl[-M_X^2(x+z-1)-y(M_X^2+s-m_0^2)+m_0^2(x+z)^2+y^2s+y(x+z)(s+m_0^2-M_Y^2)\Bigr]^{d/2-4} \\
&\quad -t\, zx\frac{d-8}{2}\Bigl[ -M_X^2(x+z-1)-y(M_X^2+s-m_0^2) 
+m_0^2(x+z)^2+y^2s+y(x+z)(s+m_0^2-M_X^2)  \Bigr]^{d/2-5}\bigg\} ~, \nonumber
\end{align}
where the external particles are put on-shell. We combine 
baryon parameters to $u:=x+z$ and $v:=x-z$ and, after 
$v$-integration, also $y+u$ and $y-u$, such that we obtain a
combination of the parameter integrals of the form~\eqref{eqn:AF}
\begin{align}
I_{MBBB}&=\frac{\Gamma(4-d/2)M_X^{d-5}}{4m_0^{d-2}(4\pi)^{d/2}\sqrt{s}(d-6)}\biggl\{
\frac{4}{K}\mathcal{D}^{3,1} -\frac{4(\mathcal{A}^{2,0}-\mathcal{D}^{2,0})}{K^2(d-4)}  
\nn
&\quad-\frac{t}{m_0^2(d-4)(d-2)}\biggl[\frac{\mathcal{A}^{1,0}  -\mathcal{D}^{1,0}}{K^4}  -\frac{d^2-6d+4}{2K^2}\mathcal{D}^{3,2}+\frac{d^3-12d^2+44d-48}{6K}\mathcal{D}^{4,3} 
+\frac{d-2}{K^3}\mathcal{D}^{2,1}\biggr]\biggr\} ~,
\end{align}
with
\beq
C_0=1-c_0^2,\quad c_0=\frac{s_{thr}+M_X^2-m_0^2}{2M_X\sqrt{s_{thr}}} ~, \quad
D_0=1-d_0^2,\quad d_0=\frac{M_X}{m_0} ~,
\eeq
where $s_{thr}=(m_0+M)^2$ with the appropriate meson mass $M$.
The $\mathcal{A}$-type integrals again demand analytic
continuation to $d=4$ as presented in Appendix~\ref{App:analytic-continuation}.  
The result of the final parameter integration is given in
Appendix~\ref{App:Integrals-results}.

\subsection{Analytic continuation of the parameter integrals}\label{App:analytic-continuation}

We here show the recursion formulas necessary for the analytic continuation
of the integrals of type $\mathcal{A}^{Z,y}$ as defined in~\eqref{eqn:AF}.
One finds
\begin{align}
\mathcal{A}^{3,2}&:=-\frac{1}{d-4}(\mathcal{A}^{2,0}+c_0)+c_0^2\mathcal{A}^{3,0}~, & 
\mathcal{A}^{3,0}&:=\frac{1}{\sqrt{C_0}}\Big(\arctan\Big(\frac{c_0}{\sqrt{C_0}}\Big)+\frac{\Gamma^2(1/2)}{2\Gamma(3-d/2)}\Big)~, \nn
\mathcal{A}^{2,0}&:=\frac{d-4}{d-3}(C_0\mathcal{A}^{3,0}+\frac{c_0}{d-4}) ~,&
\mathcal{A}^{1,0}&:=\frac{d-2}{d-1}(C_0\mathcal{A}^{2,0}+\frac{c_0}{d-2})~,\nn
\mathcal{A}^{2,1}&:=-\frac{1}{d-2}+c_0\mathcal{A}^{2,0} ~,&
\mathcal{A}^{3,1}&:=-\frac{1}{d-4}+c_0\mathcal{A}^{3,0} ~,\nn
\mathcal{A}^{2,2}&:=-\frac{1}{d-2}(\mathcal{A}^{1,0}+c_0)+c_0^2\mathcal{A}^{2,0} ~, &
\mathcal{A}^{3,3}&:=c_0\mathcal{A}^{3,2}-\frac{2}{d-4}\mathcal{A}^{2,1} ~, \nn
\mathcal{A}^{4,0}&:=\frac{1}{2C_0}\Bigl(c_0+\frac{1}{\sqrt{C_0}}\arctan\Big(\frac{c_0}{\sqrt{C_0}}\Big)+\frac{1}{\sqrt{C_0}}\frac{\Gamma(3/2)\Gamma(1/2)}{\Gamma(6-d)}\Bigr) ~, \hspace*{-5cm} \nn
\mathcal{A}^{4,3}&:=-\frac{1}{d-6}\Big(2\mathcal{A}^{3,1}+c_0\mathcal{A}^{3,0}+c_0^2-(d-6)c_0^3\mathcal{A}^{4,0}\Big) ~.\hspace*{-2cm} 
\end{align}

\subsection{Result of the threshold expansion}\label{App:Integrals-results}

Here we present the results of the expansion of the nontrivial
one-loop integrals at threshold, i.e.\ $s_{thr}=(m_0+M_X)^2$, $u_{thr}=(m_0-M_X)^2$, 
where $M_X$ corresponds to the mass of the external and $M_Y$ of the internal meson. 
Loop integrals with two meson propagators were already given in Appendix~\ref{App:thr-exp}.
We suppress the higher orders of the respective expansion parameter.

\subsubsection{Two-point functions}

\begin{align}
I_{MB}(s_{thr},M_Y)&=\frac{1}{32\pi ^2 (m_0+M_X)^2 }\bigg\{\Big(M_Y^2+2
m_0M_X+M_X^2\Big)\big(1-32\pi^2\lambda_{Y}\big) 
\nn&\qquad\qquad\qquad\qquad\qquad
+ 2\sqrt{-(M_Y^2-M_X^2) (M_Y^2-(2
m_0+M_X)^2)}\arccos\Big(\frac{M_Y^2+M_X^2+2 m_0 M_X}{2 M_Y m_0+2 M_Y
M_X}\Big)\bigg\} ~,\nn
I_{MB}(u_{thr},M_Y)&=I_{MB}(s_{thr},M_Y) \big|_{M_X\to-M_X} ~,\nn
I_{MB}(m_0^2,M_Y)&=\frac{1}{32\pi ^2m_0^2
}\bigg\{M_Y^2\big(1-32\pi^2\lambda_{Y}\big)-2M_Y \sqrt{4
m_0^2-M_Y^2}\arccos\Big(-\frac{2 m_0}{M_Y}\Big)\bigg\} ~.
\end{align}

\subsubsection{Three-point functions}

In the following we use $\epsilon:=M_X^2-(\sqrt{s}-m_0)^2$ and
\begin{align}
\mathcal{P} &:=\sqrt{4 m_0^2-M_Y^2}\arccos\Big(\frac{M_Y}{2m_0}\Big) ~, \nn
\mathcal{R} &:=\sqrt{-(M_X^2-M_Y^2) ((2 m_0+M_X)^2-M_Y^2)}\arccos\Big(\frac{M_X^2+2 m_0 M_X+M_Y^2}{2M_Y m_0+2M_YM_X}\Big)~.
\end{align}
We find
\begin{align}
I_{MBB}(m_0^2,m_0^2,M_Y^2)&=\frac{1}{32\pi ^2m_0^2
}\bigg\{-32\pi^2\lambda_{Y}-1-\frac{ 2M_Y }{\sqrt{4
m_0^2-M_Y^2}}\arccos\Big(-\frac{M_Y}{2 m_0}\Big)\bigg\} , \nn
I_{MBB}(m_0^2,s_{thr},M_Y^2)&=
\frac{1}{32 \pi ^2m_0 M_X (m_0+M_X) (m_0 (2 m_0+M_X)-M_Y^2) }\nn
&\times\bigg\{M_X (m_0 (2 m_0+M_X)-M_Y^2) (32 \pi ^2 \lambda_Y-1)+
2(m_0+M_X) M_Y\mathcal{P} - 2 m_0\mathcal{R}\nn
&\qquad-\frac{\epsilon}{36 m_0^2 M_X^2 (m_0 (2 m_0+M_X) -M_Y^2)^2}\nn
&\qquad\quad\times\bigg(96 M_X^3 \pi ^2 \lambda_Y(2 m_0^2+M_X
m_0-M_Y^2)^3 \nn
&\qquad\quad+M_X (2 m_0^2+M_X m_0-M_Y^2) \Big[4 M_X^2 m_0^4-48 M_Y^2
m_0^4+4 M_X^3 m_0^3-72 M_X M_Y^2 m_0^3\nn
&\qquad\quad+M_X^4 m_0^2+12 M_Y^4 m_0^2-16 M_X^2 M_Y^2 m_0^2+12 M_X
M_Y^4 m_0+10 M_X^3 M_Y^2 m_0+M_X^2 M_Y^4\Big]\nn
&\qquad\quad+6M_Y^3(2 m_0-M_X) (m_0+M_X)^2  (4 m_0^2+3 M_X
m_0-M_Y^2)\mathcal{P}\nn  
&\qquad\quad-6 m_0^2\Big[ m_0 M_X^4+4 m_0^2 M_X^3-3 M_Y^2 M_X^3+4 m_0^3
M_X^2\nn
&\qquad\quad+m_0 M_Y^2 M_X^2-3 M_Y^4 M_X+14 m_0^2
M_Y^2 M_X-2 m_0 M_Y^4+8 m_0^3 M_Y^2\Big]\mathcal{R}\bigg)\bigg\}~,\nn
I_{MBB}(m_0^2,u_{thr},M_Y^2)&=I_{MBB}(m_0^2,s_{thr},M_Y^2)
\big|_{M_X\to-M_X} ~.
\end{align}
	
\subsubsection{Four-point functions}

\begin{align}
I_{MBBB}(s_{thr},M_Y^2)&=\frac{1}{16 \pi ^2m_0 M_X^2  (2 m_0^2+M_X m_0-M_Y^2)^2}\nn
&\times\bigg\{M_X (2 m_0^2+M_X m_0-M_Y^2)-\frac{m_0 M_Y}{(4 m_0^2-M_Y^2)}(4
m_0^2+2 M_X m_0-M_X^2-M_Y^2)\mathcal{P}+m_0\mathcal{R}\nn
&\qquad+\frac{t}{72 m_0^2 M_X^2  (4 m_0^2-M_Y^2)^2 (2
m_0^2+M_X m_0-M_Y^2)^2} \nn
&\qquad\quad\times\bigg(    M_X (4 m_0^2-M_Y^2) (m_0 (2 m_0+M_X)-M_Y^2)
\Big[(-6 m_0^2+3 M_X m_0-2 M_X^2) M_Y^6\nn
&\qquad\quad+m_0 (48 m_0^3+6 M_X m_0^2+19 M_X^2 m_0+7 M_X^3)
M_Y^4-m_0^2 (96 m_0^4+72 M_X m_0^3\nn
&\qquad\quad+88 M_X^2 m_0^2+48 M_X^3 m_0+11 M_X^4) M_Y^2+32 m_0^4
M_X^2 (2 m_0+M_X)^2\Big]\nn
&\qquad\quad-6m_0^3(M_Y^2-M_X^2) ((2 m_0+M_X)^2-M_Y^2)(M_Y^2-4
m_0^2)^2\mathcal{R}\nn
&\qquad\quad+ 6 M_Ym_0^3\big(4 m_0^2+2 M_X
m_0-M_X^2-M_Y^2\big)\Big[M_Y^6+16 m_0^4 M_Y^2-6 m_0^2 M_X^2 (2 m_0+M_X)^2\nn
&\qquad\quad+(16 M_X m_0^3+20 M_X^2
m_0^2+8 M_X^3 m_0+M_X^4)-4 (2
m_0^2+M_X m_0+M_X^2) M_Y^4\Big]\mathcal{P}\bigg)\bigg\}~,\nn
I^C_{MBBB}(u_{thr},M_Y^2)&=I_{MBBB}(s_{thr},M_Y^2)
\big|_{M_X\to-M_X} ~.
\end{align}

\section{Result of the SU(2) expansion of the loop integrals}\label{App:MatIntr}

Here we present the results of the double-scale expansion of the scalar loop integrals as described in the main text.
The integrals are expanded up to the order required by the calculation in the main part. 
We use the abbreviations $\XX:=s-m_0^2$, $\YY:=u-m_0^2$.

\subsection{Two-point functions}
\vskip -1mm
\begin{align}
I_{MB}(s,M^2_Y) =
\frac{1}{32\pi^2m_0^2}&\bigg\{\frac{(10m_0^4-6M_Y^2m_0^2+M_Y^4)\XX^2}{2m_0^4(4m_0^2-M_Y^2)}+32\pi^2\lambda_Y\Big(-M_Y^2+\frac{(m_0^2-M_Y^2)\XX^2
}{m_0^4}-\frac{(m_0^2-M_Y^2)\XX }{m_0^2}\Big)\nn
&-\XX+M_Y^2+\frac{2\arccos\left(-\frac{M_Y}{2m_0}\right)}{m_0^4(4m_0^2-M_Y^2)^{3/2}M_Y}\bigg((2m_0^6-12M_Y^2m_0^4+7M_Y^4m_0^2-M_Y^6)\XX^2
\nn
&-m_0^4M_Y^2(4m_0^2-M_Y^2)^2+m_0^2M_Y^2(3m_0^2-M_Y^2)(4m_0^2-M_Y^2)\XX\bigg)\bigg\}
~,\nn
I_{MB}(u,M^2_Y)=I_{MB}(s,M_Y&)\Big|_{\XX\rightarrow\YY} ~, \nn
I_{MB}(m_0^2,M^2_Y)
=\frac{1}{32\pi^2m_0^2}&\bigg\{M_Y^2(1-32\pi^2\lambda_Y)-2M_Y\sqrt{4m_0^2-M_Y^2}\arccos\Big(-\frac{M_Y}{2
m_0}\Big)\bigg\}~.
\end{align}

\subsection{Three-point functions}
\vskip -1mm
\begin{align}
I_{MMB}(t,M^2_Y,M^2_Y)
=\frac{1}{32\pi^2m_0^2}&\bigg\{1+32\pi^2\lambda_Y+\frac{4m_0^2-2M_Y^2}{M_Y\sqrt{4m_0^2-M_Y^2}}\arccos\Big(-\frac{M_Y}{2
m_0}\Big)\bigg\} ~,\nn
I_{MBB}(m_0^2,s,M_Y^2)
=\frac{1}{32\pi^2m_0^2}&\bigg\{32\pi^2\lambda_Y\Big(-\frac{
M_\pi^2}{6m_0^2}-1+\frac{\XX}{2m_0^2}-\frac{2\XX^2 }{3m_0^4}\Big)
-\frac{(2m_0^2-M_Y^2)M_\pi^2}{3m_0^2(4m_0^2-M_Y^2)}-1\nn
&+\frac{(16m_0^6-70M_Y^2m_0^4+38M_Y^4m_0^2-5M_Y^6)\XX^2}{3m_0^4M_Y^2(4m_0^2-M_Y^2)^2}+\frac{(3m_0^2-M_Y^2)\XX}{m_0^2(4m_0^2-M_Y^2)}\nn
&+\frac{\arccos\left(-\frac{M_Y}{2m_0}\right)}{3m_0^4M_Y(4m_0^2-M_Y^2)^{5/2}}
\bigg[-4\big(18m_0^6-30M_Y^2m_0^4+10M_Y^4m_0^2-M_Y^6\big)\XX^2 \nn
&+6m_0^4M_Y^2(4m_0^2-M_Y^2)^2 +M_Y^2M_\pi^2m_0^2(6m_0^2-M_Y^2)(4m_0^2-M_Y^2)\nn
&+3(4m_0^4-6M_Y^2m_0^2+M_Y^4)(4m_0^2-M_Y^2)\XX\bigg]\bigg\} ~,\nn
I_{MBB}(m_0^2,u,M_Y^2) =I_{MBB}(M_Y&)\Big|_{\XX\rightarrow\YY} ~, \nn
I_{MBB}(m_0^2,m_0^2,M_Y^2)
=\frac{1}{96\pi^2m_0^2}&\bigg\{-96\pi^2\lambda_Y\Big(1+\frac{t}{6m_0^2}\Big)
-\frac{(2m_0^2-M_Y^2)t}{m_0^2(4m_0^2-M_Y^2)}-3 
\nn&
+\frac{\arccos\left(-\frac{M_Y}{2m_0}\right)}{m_0^2(4m_0^2-M_Y^2)^{3/2}}\Big(M_Y(6m_0^2-M_Y^2)t+6M_Ym_0^2(4m_0^2-M_Y^2)^2\Big)\bigg\}
~.
\end{align}

\subsection{Four-point functions}
\vskip -1mm
\begin{align}
I_{MBBB}(s,M_Y^2)&=\frac{1}{32\pi^2m_0^2(4m_0^2-M_Y^2)}\bigg\{1+\frac{4m_0^2}{M_Y\sqrt{4m_0^2-M_Y^2}}\arccos\Big(-\frac{M_Y}{2
m_0}\Big)\bigg\}
~,\nn
I_{MBBB}(u,M_Y^2) &=I_{MBBB}(s,M_Y^2) ~.
\end{align}

\end{widetext}



\end{document}